\documentclass[onecolumn,showpacs,preprintnumbers,floatfix,amsmath,amssymb,aps]{revtex4}
\usepackage{bm}
\usepackage{color}

\usepackage{graphicx,subfig,epsfig,epsf,color}
\usepackage[utf8x]{inputenc}
\usepackage{amssymb,amsmath,appendix}
\usepackage{slashed}
\usepackage{array}
\usepackage{hyperref}
\usepackage{ulem}

\begin{document}
\title{Speed of Sound in Hybrid Stars and the Role of Bag Pressure in the Emergence of Special Points on $M-R$ Variation of Hybrid Stars}

\author{Suman Pal$^{1,2}$, Soumen Podder$^1$, Debashree Sen$^1$, and Gargi Chaudhuri$^{1,2}$}

\address{$^1$Physics Group, Variable Energy Cyclotron Centre, 1/AF Bidhan Nagar, Kolkata 700064, India}
\address{$^2$Homi Bhabha National Institute, Training School Complex, Anushakti Nagar, Mumbai 400085, India}

\email{suman.p@vecc.gov.in, s.podder@vecc.gov.in, debashreesen88@gmail.com, gargi@vecc.gov.in}

\date{\today}




\begin{abstract}

We compute the hybrid star (HS) properties with the help of Maxwell construction. For the purpose we choose a fixed hadronic model and four different forms of MIT bag model for the quark phase. We investigate thoroughly the effects of the different parameters of the bag model on the speed of sound in HS matter and the structural properties of HSs in the light of the various recent constraints on them from astrophysical observations. We also examine the importance of each parameter involved in these four forms of bag model in the context appearance of special points (SPs) in the mass-radius ($M-R$) variation of HSs. We find that among all these parameters the bag pressure play the most significant role in the emergence of the SPs in the $M-R$ dependence of HSs. 

\end{abstract}




\maketitle



\section{Introduction}
\label{Intro}

 The phenomenon of hadron-quark phase transition is one of the most interesting current topics of research which finds its application in various interesting contexts like the heavy-ion collision physics, supernova explosions and binary neutron star mergers (BNSMs) etc. Quantum Chromodynamics (QCD) calculations help us to picturize the QCD phase diagram along the chemical potential (density) and temperature axes and speculate the existence of quark-gluon plasma (QGP) at high temperature (relevant to the early stages of the universe) and at high density (relevant to compact star cores). While the high temperature - low baryon density regime of the QCD phase diagram is accessible to the heavy-ion collision experiments to a certain extent, the low (negligible) temperature - high density regime conditions are highly challenging to attain from experimental perspectives. Such ideal conditions are characteristics of the core of neutron/compact stars. In the present era of the BNSM detection especially GW170817, the phenomena of phase transition has gained special attention and interest. It has been suggested that if the post-merger phase of the GW170817 event can be detected in future then the data analysis of both the inspiraling and post-merger phases can provide further insight into the compact star properties and may indicate the possibility of first-order phase transition in the BNSM \cite{BNSM}. At present unfortunately the lack of concrete experimental evidence makes the understanding of matter and its composition, its interactions and the equation of state (EoS) at such high density quite inconclusive. Theoretical studies suggest that such conditions can support the formation of exotic matter like the hyperons, delta baryons and boson condensates etc. \cite{Glendenning,Weissenborn12,Logoteta,Sen6,Sen7}. Also at such high densities the asymptotic freedom of QCD suggest a possible first order phase transition from hadronic matter to quark matter \cite{Glendenning,Blaschke2,Weissenborn2011,Bhattacharya,Gomes2019,Han,Ferreira2020, Most2,Zha,Khanmohamadi,Xia2,Maslov,Montana,Christian,Contrera,Lugones2021,Bozzola, Liu2022,cs3,Marczenko,Aguirre,Blaschke,Ivanytskyi,Alvarez-Castillo1,Schertler, Blaschke2015,Alvarez-Castillo,Cierniak,Sharifi,Ayriyan,Jakobus,Espino,Kaltenborn, Yudin,Cierniak2,Ju,cs_Gibbs,Agrawal,Lopes_Hybrid,Sen6,Sen7,Sen8,Sen9}. This leads to the possible formation of hybrid stars (HSs). Theoretical formulation of compact star EoS is thus related to the particular consideration of its composition and the interactions. Therefore compact star EoS is subjected to a lot of uncertainties. Fortunately, compact star EoS is constrained to certain extent by some astrophysical and observational results such as those on their maximum mass obtained from high mass pulsars like PSR J0348+0432 \cite{Ant} and PSR J0740+6620 \cite{Fonseca2021}. Recently, NICER experiment also constrained the radii of PSR J0740+6620 \cite{Miller2021,Riley2021} and PSR J0030+0451 \cite{Riley2019,Miller2019}. The GW170817 observational data also set limits on the dimensionless tidal deformability and radius of of a 1.4 $M_{\odot}$ compact star \cite{GW170817}.

 In general the hadron-quark phase transition is achieved in HSs  with the help of Gibbs and/or Maxwell constructions depending on the value of surface tension at the hadron-quark boundary \cite{Maruyama}. The former is based on global charge neutrality condition, characterized by the formation of mixed phase \cite{Glendenning,Bhattacharya,Sen6} while in the later the local charge neutrality condition is considered and considerable density jump is noted  \cite{Bhattacharya,Gomes2019,Han,Ferreira2020,Khanmohamadi,Sen7}. When the surface tension at the boundary becomes quite high, the Maxwell construction is favored since under such conditions the mixed phase (with Gibbs construction) becomes unstable. In the present we adopt Maxwell construction by assuming the surface tension at the hadron-quark phase boundary to be high enough and obtain the properties of HSs. The study of phase transition in HSs have led to several interesting results and possibilities. One of them is the variation of speed of sound ($C_S$) in HS matter and it has been also seen that its variation in HS matter often follows a uphill-downhill nature along density. Moreover, strong first order phase transition often leads to the drastic surpassing of the conformal limit of $C_S^2=1/3$ and close to the causality limit of $C_S^2<1$  \cite{cs3,Marczenko,Reed,Tan,Aguirre,Blaschke,Ivanytskyi,Alvarez-Castillo1,Sen6}. Thus in the present work we study the variation of speed of sound under certain conditions of phase transition using different quark models.

 Another interesting feature of HSs is that irrespective of the type of construction or the hadronic/quark model used, there may be formation of “third family” of compact stars \cite{Glendenning,Maslov,Schertler,Blaschke2015,Alvarez-Castillo,Alvarez-Castillo1,Blaschke,Cierniak,Sharifi,Ayriyan,Jakobus,Espino,Wang, Sen9} and twin star configurations under certain circumstances of strong first order phase transitions mainly with considerable density jumps. The mass-radius ($M-R$) variation of the HSs in such cases is characterized by non-identical branches with two distinct maximas at two different radii. This may often lead to the appearance of twin stars which are actually two separate points on the $M-R$ plot of the HSs with same mass but different radii. Of them, one is generally located on the regular neutron (hadronic) star branch or the first stable branch while the other is a HS lying on the second stable branch. The two branches are often disconnected from each other by an instable region \cite{Alvarez-Castillo,Alvarez-Castillo1,Kaltenborn,Ivanytskyi,Christian,Sen9}. However, the location of the twins depends on the transition density of the HSs. So the twins with nearly identical mass can both also be located in the second stable branch \cite{Lyra}. Twin stars are broadly into four categories depending on their location in the $M-R$ plot \cite{Montana,Christian,Sharifi}. Third family of compact stars may also show a special feature of special points (SPs). A SP ($M_{SP}$, $R_{SP}$) on the $M-R$ plot is a narrow region through which all the HS solutions pass irrespective of the different transition densities for different values of bag pressure, the model or the type of construction adopted to achieve hadron-quark phase transition \cite{Cierniak,Yudin,Cierniak2,Kaltenborn,Blaschke3, Blaschke,Ayriyan,Ivanytskyi,Wang,Sen9}. The appearance of SPs on the mass-radius diagram of HSs is an effect of variation of certain parameters related to different quark models \cite{Yudin,Wang,Cierniak,Cierniak2,Kaltenborn,Blaschke3,Ivanytskyi,Blaschke,Alvarez-Castillo1}. The existence of such SPs can thus be treated as universal property of HS models and they serve as a remarkable tool to interpret the multi-messenger data as signals for the possible existence of HSs \cite{Cierniak,Cierniak2,Sen9}.
 
 Recently, we studied HS properties obtained with different hadronic models and the MIT bag model with density dependent bag pressure in a Gaussian distribution form \cite{Burgio1,Burgio2} with special emphasis on the formation of third family of compact stars and the emergence of SPs \cite{Sen9}. We showed that the mass corresponding to SP ($M_{SP}$) and the maximum mass ($M_{max}$) of the HSs follow a nearly linear (fitted) relationship where the slope is independent of the value of bag pressure. Since this $M_{SP}-M_{max}$ dependence of the HSs is found to be consistent with any hadronic EoS chosen to obtain the hybrid EoS, such relations are treated as universal relations in the context of formation of SPs. In the present work we now study the HS properties with a fixed hadronic model and four different forms of the MIT bag model. Within the framework of the MIT Bag model, we intend to examine thoroughly the different parameters of the bag model involved in these forms and indicate specifically the ones responsible for formation of SPs on the $M-R$ solution of HSs. For the purpose we choose the well-known relativistic mean-field (RMF) hadronic model BSR2 \cite{bsr} and the four different forms of MIT bag model which includes the modified bag model including the strong repulsive interaction \cite{Farhi,Glendenning} and the vector bag (vBag) model \cite{Lopes,Vivek,SenGuha}. Considering the fact that the quarks acquire asymptotic degree of freedom at high densities relevant to HS cores, in the present work we also consider the density dependent of the bag model following a Gaussian distribution form of bag pressure \cite{Burgio1,Burgio2} since this form includes the concept of asymptotic freedom of the quarks. Few works \cite{Alvarez-Castillo1,Contrera} considered the dependence of bag pressure with respect to chemical potential in a hyperbolic form. Motivated by such works we also considered density dependence of bag pressure in hyperbolic form. Such a form was also adopted by \cite{Masuda} to construct the mixed phase of HSs. With these four different forms of bag model we compute the hybrid EoS and study the HS structural properties in the light of various astrophysical constraints. We also examine the importance of each parameter involved in these four forms of bag model on the HS properties with special emphasis to their connection with the appearance of SPs in the $M-R$ variation of HSs. We adopt Maxwell construction for phase transition and for the pure hadronic phase we consider the BSR2 model \cite{bsr}. It is well-known that the presence of heavier baryons like the hyperons and delta baryons soften the EoS and reduce the maximum mass of the NSs \cite{Glendenning,Weissenborn12,Logoteta, Sen6,Sen7} and there is still a lot of uncertainty pertaining to the hyperon couplings \cite{Logoteta}. Moreover, the conditions of temperature, density and iso-spin asymmetry in compact star cores are not well-known which broadly influence the threshold for the appearance or disappearance  of these exotics \cite{Blaschke2015}. It is also suggested that in case of HSs the threshold density of appearance of hyperons and quarks may often be very close or even overlapping \cite{Blaschke2015}. Therefore, for all these reasons and similar to \cite{Ju,Liu2022,Agrawal,Sen8,Sen9} we do not consider the presence of hyperons or delta baryons in the hadronic phase of HSs of the present work.
 
 This paper is organized as follows. In the next section \ref{Formalism}, we address the hadronic model adopted (Section \ref{Hadronic_model}). In Section \ref{Quark phase}, the main features of the four different forms of the bag model are highlighted. We then present our results and corresponding discussions in section \ref{Results}. We summarize and conclude in the final section \ref{Conclusion} of the paper.
  

\section{Formalism}
\label{Formalism}

\subsection{Pure Hadronic Phase}
\label{Hadronic_model}

 For the pure hadronic phase, we employ the well-known RMF model BSR2 \cite{bsr}. The saturation properties of this model is in reasonable agreement with the different experimental and empirical data. In table \ref{tab:1} we list the saturation properties like the saturation density ($\rho_0$), binding energy per particle ($e_0$), nuclear incompressibility ($K_0$), symmetry energy coefficient ($J_0$) and the slope parameter ($L_0$) of the chosen hadronic model.

\begin{table}[!ht]
\caption{The nuclear matter properties at saturation density $\rho_{_{0}}$ for the chosen hadronic model.}
\setlength{\tabcolsep}{12.0pt}
\begin{tabular}{cccccc}
\hline
\hline
$\rho_{0}$ & $e_{0}$ & $K_{0}$ & $J_{0}$ & $L_{0}$ \\
$({\rm fm}^{-3})$ & (MeV) & (MeV) & (MeV) & (MeV) \\ 
\hline
0.149 & $-$16.03 & 240.0 & 31.4 & 62.2   \\
\hline
\hline
\end{tabular}
\label{tab:1}
\end{table}

 The symmetry energy coefficient ($J_0$) and the slope parameter ($L_0$) of the chosen hadronic model are quite consistent with the recent findings of \cite{Reed} obtained from the correlation between them and the neutron skin thickness of $^{208}\rm{Pb}$ ($R^{208}_{skin}$) as measured by the PREX-II experiment. The binding energy per particle ($e_0$) and the saturation density ($\rho_0$) also satisfy the experimental constraints \cite{Dutra}. The nuclear incompressibility ($K_0$) is also in good agreement with that prescribed from the experimental finding of \cite{K}. In addition to the constraints on nuclear saturation properties, the above parameterization also satisfies the data obtained from finite nuclei experiments \cite{FinNuc}. The chosen hadronic model has been well adopted in literature, even in recent works, to determine the properties of neutron/hybrid stars.

 As mentioned in section \ref{Intro} we do not include the hyperons or the delta baryons in the hadronic sector and in the present work we consider $\beta$ stable hadronic matter consisting of the nucleons, electrons and muons as the composition of the hadronic phase.

\subsection{Pure Quark Phase and Hadron-Quark Phase Transition}
\label{Quark phase}

 We adopt four different forms of the MIT Bag model with u, d and s quarks along with the electrons to describe the pure quark phase. The original bag model was formulated based on the hypothesis that the unpaired quarks are constrained within a hypothetical region called `Bag', characterized by specific bag pressure that determines the strength of quark interaction \cite{Chodos}. This bag pressure signifies the difference in energy density between the perturbative vacuum and the true vacuum \cite{Burgio1,Burgio2}. The value of $B$ is still inconclusive and it is often taken as free parameter that plays an important role in determining the properties of the HSs. Over the years this original and simplistic form of the bag model was modified rigorously into further several other realistic and sophisticated forms. In the present we choose four such different forms of the MIT bag model and employ each to compute the HS properties. The four types of bag model adopted in the present work are as follows.
 
\subsubsection{Modified Bag Model with Strong Repulsive Interaction}
\label{Formalism_alpha}
 
 Strong repulsive interaction between the quarks was introduced by \cite{Farhi} in terms of a repulsive interaction coupling parameter $\alpha$. The thermodynamic potential \cite{Farhi,Glendenning} to first order in strong interaction is given as 

\begin{eqnarray}
\Omega_f=-\frac{\gamma_f}{24\pi^2}\Bigg[\mu_f \sqrt{\mu_f^2 - m_f^2} \Big(\mu_f^2 - \frac{5}{2}m_f^2\Big) + \frac{3}{2}m_f^4~ ln\Bigg(\frac{\mu_f + \sqrt{\mu_f^2 - m_f^2}}{m_f}\Bigg) \nonumber \\ - \frac{2\alpha}{\pi}\Bigg\lbrace3\Bigg(\mu_f \sqrt{\mu_f^2 - m_f^2} - m_f^2~ ln\frac{\mu_f + \sqrt{\mu_f^2 - m_f^2}}{m_f}\Bigg)^2 -2 \Big(\mu_f^2 - m_f^2\Big)^2  - 3m_f^4~ ln^2 \Big(\frac{m_f}{\mu_f}\Big)\Bigg\rbrace\Bigg]
\label{omg_alp}
\end{eqnarray} 

 where, $f$ = u, d and s are the quark flavors. $m_f$ and $\mu_f$ are the mass and the chemical potential, respectively, of individual quark flavor and $\gamma_f$ is the spin degeneracy factor. In \cite{Farhi} the masses of u and d quarks is totally neglected with respect to that of s quark and therefore in \cite{Farhi} the above form in equation \ref{omg_alp} is only used for the s quark. However, in the present work we have adopted the form presented in \cite{Glendenning} where $m_u$ and $m_d$ are not neglected. We therefore consider small but finite mass of both u and d quarks as $m_u$=2.16 MeV and $m_d$=4.67 MeV \cite{PDG}. However, we do not consider the contribution from the term due to renormalization as considered in both \cite{Farhi,Glendenning}. In case of electrons, whose mass is neglected, the thermdynamic potential $\Omega_e$ can be calculated by replacing $m_f$ by $m_e=0$ and $\gamma_f$ by $\gamma_e$ in the first term of equation \ref{omg_alp} that do not involve the repulsive coefficient $\alpha$. Therefore $\Omega_e=\frac{-\mu_e^4}{12 \pi^2}$ and the total thermodynamic potential of the quark phase including quarks and electrons becomes $\Omega = \Omega_f + \Omega_e + B$. The quark and baryon density can be obtained using equation \ref{omg_alp} while formulation of the EoS (energy density $\varepsilon$ and pressure $P$) involves the expression of $\Omega$ \cite{Glendenning}. Thus among these quantities the bag constant $B$ contribute only to the EoS \cite{Farhi,Glendenning}.

\subsubsection{Vector Bag (vBag) Model}
\label{Formalism_vBag}

The repulsive effect of quark interaction was also included by introducing the vector meson as mediator (vBag model) \cite{Lopes,Vivek,SenGuha}.

\begin{eqnarray}
\mathcal{L} = \sum_f \Bigg[\overline{\psi}_f \bigg\lbrace \gamma^{\mu} \big(i \partial_{\mu} - g_{qqV}V_{\mu}\big) - m_f \bigg\rbrace \psi_f - B \Bigg] \Theta (\overline{\psi}_f \psi_f) \\ \nonumber + \frac{1}{2} m_V^2 V_{\mu} V^{\mu} - \frac{1}{4} V_{\mu\nu} V^{\mu\nu} + b_4 \frac{(g^2V_{\mu} V^{\mu})^2}{4} + \overline{\psi}_l \big(i \gamma_{\mu} \partial^{\mu} - m_l\big) \psi_l
\label{Lagrangian_vBag}
\end{eqnarray} 

where, $f$ = u, d and s and the lepton $l$=e, the electrons. $B$ is the Bag constant and the Heaviside function $\Theta$=1 inside the bag. The scaled couplings are defined as $X_V=g_{ssV}/g_{uuV}$ and $G_V=(g_{uuV}/m_V)^2$. So $G_V=$0 reduces to the original form of the MIT Bag model without interactions. The self-interaction of the vector $\omega$ field is introduced via its quartic contribution in terms of a parameter $b_4$ that regulates the increment/decrement of the vacuum expectation value ($V_0$) of the $\omega$ field and $g=g_{uuV}$. This correction term also mimics the Dirac sea contribution of the quarks. The quark EoS for this form of the bag model can be obtained from equation \ref{Lagrangian_vBag} \cite{Lopes,Vivek}.

\subsubsection{Density Dependent Bag Model with Gaussian Form}
\label{Formalism_GDDBag}

 At high densities, relevant to NS/HS cores, the quarks gain asymptotic degrees of freedom \cite{Burgio1,Burgio2}. This indicates that the bag pressure is justifiably density dependent rather than being a constant. Therefore in the present work we consider the density dependence of the bag pressure $B(\rho)$ as the third form of bag model. Here $B(\rho)$ follows a Gaussian distribution form \cite{Burgio1,Burgio2,Sen7,Sen9} given as

\begin{eqnarray}
B(\rho) = B_{as} + (B_0 - B_{as})~ \rm{exp}~ [-\beta(\rho/\rho_0)^2]
\label{B_GDDBag}
\end{eqnarray}

where, $B_0$ and $B_{as}$ are the values attained by $B(\rho)$ at $\rho=0$ and asymptotic densities, respectively. $\beta$ controls the decrease of $B(\rho)$ with the increase of density. This form for the density dependence of the bag pressure thus involves the notion of the asymptotic behavior of the quarks at high densities. In the present work we consider the second term of RHS of equation \ref{B_GDDBag} as

\begin{eqnarray}
\Delta B=B_0 - B_{as}
\label{eq_delB}
\end{eqnarray}

 The EoS is the same obtained in the simple original form of the MIT Bag model without interaction \cite{Glendenning,Sen7,Sen9}.

\subsubsection{Density Dependent Bag Model with Hyperbolic Form}
\label{Formalism_HDDBag}

The hyperbolic form was also adopted by \cite{Masuda} to construct the mixed phase of HSs. We apply the same form to invoke density dependence of bag pressure as

\begin{eqnarray}
B(\rho) = B_0*f(\rho)
\label{B_HDDBag}
\end{eqnarray}

where,

\begin{eqnarray}
f(\rho) = \frac{1}{2} \Bigg[1 - tanh\Big(\frac{\rho-\bar{\rho}}{\Gamma_{\rho}}\Big)\Bigg]
\end{eqnarray}
 
 Such a form was also adopted by \cite{Alvarez-Castillo1,Contrera} but in terms of chemical potential. Here $B_0$, $\bar{\rho}$ and $\Gamma_{\rho}$ are free parameters. The variation of $B(\rho)$ with respect to $\rho$ follows the same type of curve in case of both Gaussian (equation \ref{B_GDDBag}) and hyperbolic (equation \ref{B_HDDBag}) forms i.e, $B(\rho)$ saturates at a particular value of $\rho$ which essentially signifies the asymptotic value of $B(\rho)$. Hence the two forms of density dependence of $B(\rho)$ are almost of same nature and carry the notion of quarks acquiring asymptotic freedom at a particular density value.

\vspace{0.5cm}

 Along with the quarks, we have also considered the contribution of electrons in the last two quark models similar to that in the first two quark models. In case of each quark model and the hadronic model the conditions of charge neutrality and the chemical potential equilibrium are imposed individually \cite{Glendenning}. Phase transition from hadronic to quark phase is obtained with Maxwell construction when the the pressure and baryon chemical potential of the individual charge neutral phases are equal \cite{Bhattacharya,Gomes2019,Han,Ferreira2020,Sen7,Sen8,Sen9}. We compute the hybrid EoS for different values of the various parameters involved in the different forms of the quark models considered. For the outer crust, the Baym-Pethick-Sutherland (BPS) EoS \cite{Baym71} is adopted upto the neutron drip density \cite{Glendenning} after which the inner crust follows which is described by the EoS that include the pasta phases in $\beta$ equilibrium condition in the form of droplets, rods and slabs structures \cite{Grill14,Alam2016}. Consequently, with the obtained hybrid EoS, we compute the speed of sound in HS matter following the equation
 
\begin{eqnarray}
C_S^2=\frac{dP}{d\varepsilon}
\end{eqnarray} 
 
 With the obtained hybrid EoS, we next proceed to study the structural properties of the HSs in static conditions like the gravitational mass ($M$) and the radius ($R$) of the HSs by integrating the Tolman-Oppenheimer-Volkoff (TOV) equations \cite{tov} based on the hydrostatic equilibrium between gravity and the internal pressure of the star. The dimensionless tidal deformability ($\Lambda$) is obtained in terms of the mass, radius and the tidal love number ($k_2$) following \cite{Hinderer}. Since we have adopted Maxwell construction, it is expected that the hybrid EoS will be characterized by jump in energy density. Therefore at the hadron-quark interface we implement the correction in calculating the second love number as suggested by \cite{k2_corr}.


\section{Results}
\label{Results}

 We study the effects of variation of the different parameters of the quark models on HS properties. We vary one parameter at a time keeping the others fixed for a particular quark model. From the first quark model viz. the modified bag model with strong repulsive interaction we vary the parameters $B$, $\alpha$ and $m_s$. As the present mass of s quark is known to be 93.4$^{+8.6}_{-3.4}$ MeV \cite{PDG}, we vary $m_s$ close to this range. From the second quark model i.e, the vBag model, we vary the parameters $B$, $G_V$ and $X_V$. After rigorous check we found that for $b_4\neq$0, the constraint on $\Lambda_{1.4}$ from GW170817 \cite{GW170817} is not satisfied for any combination of ($X_V,G_V,B$). This is also consistent with the results of \cite{Lopes_Hybrid}. Therefore in the present work we show the results for variation of each parameter among $X_V,G_V,B$ by fixing $b_4$=0. In case of pure quark stars the value $B$ corresponding to the value of $G_V$ is essential for a fixed value of $X_V$ in order to ensure the stability of the star following Bodmer-Witten conjecture \cite{Lopes,SenGuha}. However, in case of HSs, there is no such stability condition concerning the values of ($G_V,B,X_V$) that needs to be fulfilled. Therefore, in the present work we vary these three parameters as free ones in order to obtain the combinations of ($G_V,B,X_V$) for HS configurations that satisfy the various present day astrophysical constraints on the structural properties of compact stars. Each of the three parameters is varied freely keeping the other two constant in order to check their individual role on the emergence of the SPs. In case of the quark model with density dependent bag pressure in a Gaussian form, we vary $B_{as}$, $\Delta$ and $\beta$. Finally, in case of the quark model with density dependent bag pressure in a hyperbolic form, we vary $B_0$, $\bar{\rho}$ and $\Gamma_{\rho}$. In case of the first two forms described in sections \ref{Formalism_alpha} and \ref{Formalism_vBag} the bag pressure is treated as constant ($B$) while in the last two (\ref{Formalism_GDDBag} and \ref{Formalism_HDDBag}) it is taken to be density dependent ($B(\rho)$) in Gaussian and hyperbolic forms, respectively. Since the Gaussian and hyperbolic forms of have similar nature, we show the variation of speed of sound in HS matter only for the first three models.

\subsection{Speed of Sound in Hybrid Stars with different Quark Models}

  We first study the variation of speed of sound in HS matter for the different parameters involved in the first three quark models. The corresponding results are discussed briefly. Since the Gaussian and hyperbolic forms of density dependence of bag pressure have similar nature, we show the variation of speed of sound in HS matter only for the first three quark models discussed in the previous section \ref{Formalism}. The nature of variation of $C_S$ in HS matter is often dependent on the type of construction chosen to achieve phase transition. Since we consider Maxwell construction, the speed of sound in HS matter in the present work is expected to drop drastically to zero in the transition region as seen from \cite{Blaschke,Ivanytskyi,Alvarez-Castillo1,Sen7} unlike the case of Gibbs construction where speed of sound in HS matter peaks in the mixed phase region \cite{cs3,cs_Gibbs,Tan,Sen6}.
  
\begin{figure}[!ht]
\centering
\subfloat[]{\includegraphics[width=0.33\textwidth]{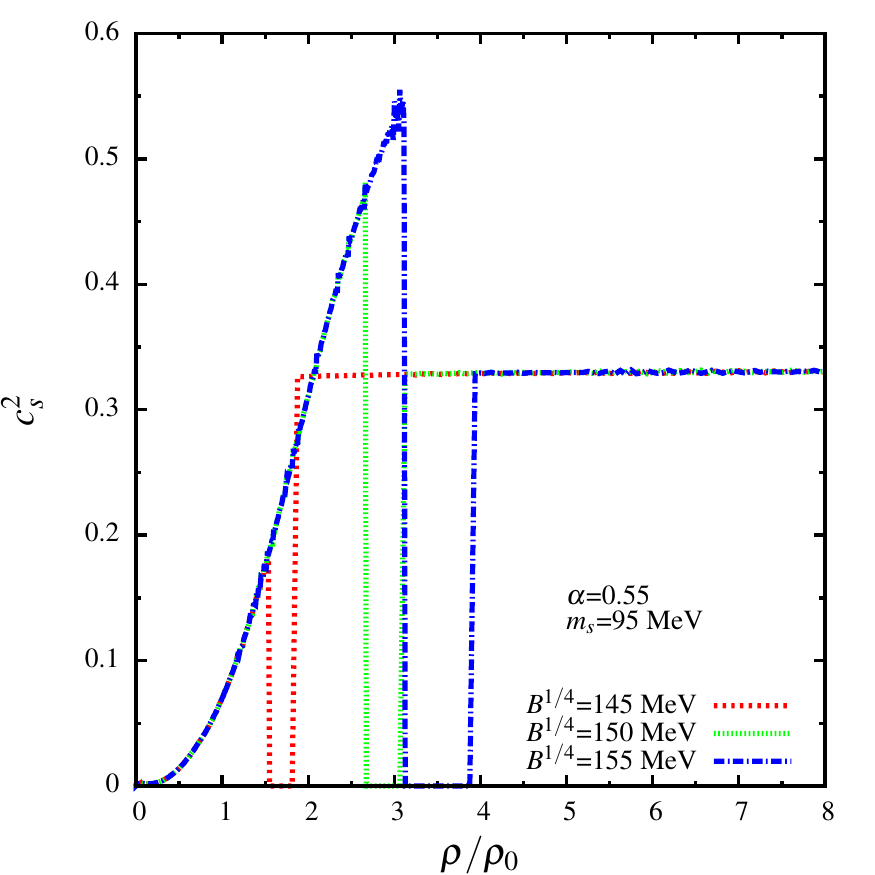}\protect\label{cs_B}}
\hfill
\subfloat[]{\includegraphics[width=0.33\textwidth]{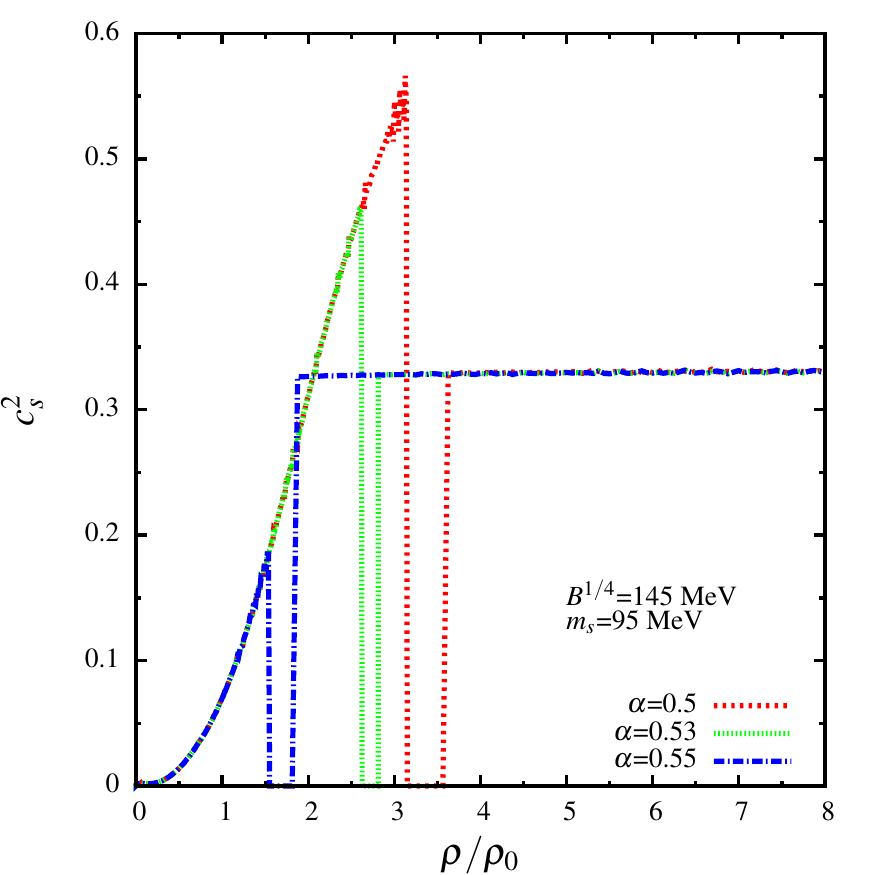}\protect\label{cs_alpha}}
\hfill
\subfloat[]{\includegraphics[width=0.33\textwidth]{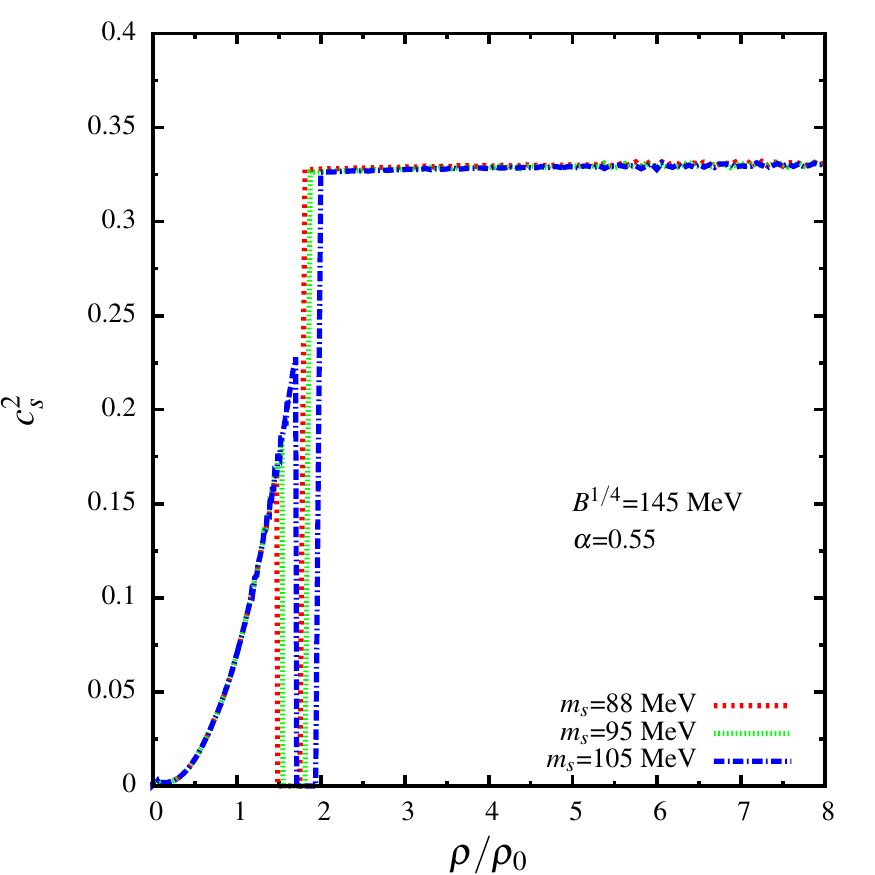}\protect\label{cs_ms}}
\caption{\it Variation of speed of sound with density of hybrid star with modified bag model for (a) different $B$ and fixed $\alpha$ and $m_s$, (b) different $\alpha$ and fixed $B$ and $m_s$, and (c) different $m_s$ and fixed $B$ and $\alpha$.}
\label{cs_mBag}
\end{figure}

 Considering the modified bag model with strong repulsive interaction, we find from figure \ref{cs_mBag} that substantial change in $C_S$ occurs due to small change in both $\alpha$ and $B$ while the variation of $m_s$ shows very feeble change in $C_S$. As expected higher value of $B$ and a lower value of $\alpha$ lead to comparatively delayed transition and hence a higher value of $C_S$. Thus in this model the speed of sound in HS is more sensitive to $\alpha$ and $B$ than $m_s$. The peak of $C_S$ is noticed just before transition in the pure hadronic phase in case of higher values of both $B$ and $\alpha$ as seen from figures \ref{cs_B} and \ref{cs_alpha} while for the lowest values of $B$ and $\alpha$ it is seen that $C_S$ peaks in the pure quark phase just like for the variation of $m_s$ as seen from \ref{cs_ms}. This implies that the transition density plays an important role in determining the location of the peak of $C_S$, whether it should lie in the hadronic or quark phase. Delayed transition tends to locate the peak of $C_S$ in the hadronic phase. For the variation of $B$ and $\alpha$, the maximum value of $C_S^2$ is very close to or above the conformal limit ($C_S^2=$0.33) but quite less than the causality limit ($C_S^2=$1). Also for this model, the speed of sound show no perceptible change in the pure quark phase for the variation of all the three parameters. The maximum value (0.57) of the peak of $C_S$ in the HS matter with this quark model is obtained for the minimum value of $\alpha$ in the hadronic phase as seen from figure \ref{cs_alpha}.

\begin{figure}[!ht]
\centering
\subfloat[]{\includegraphics[width=0.33\textwidth]{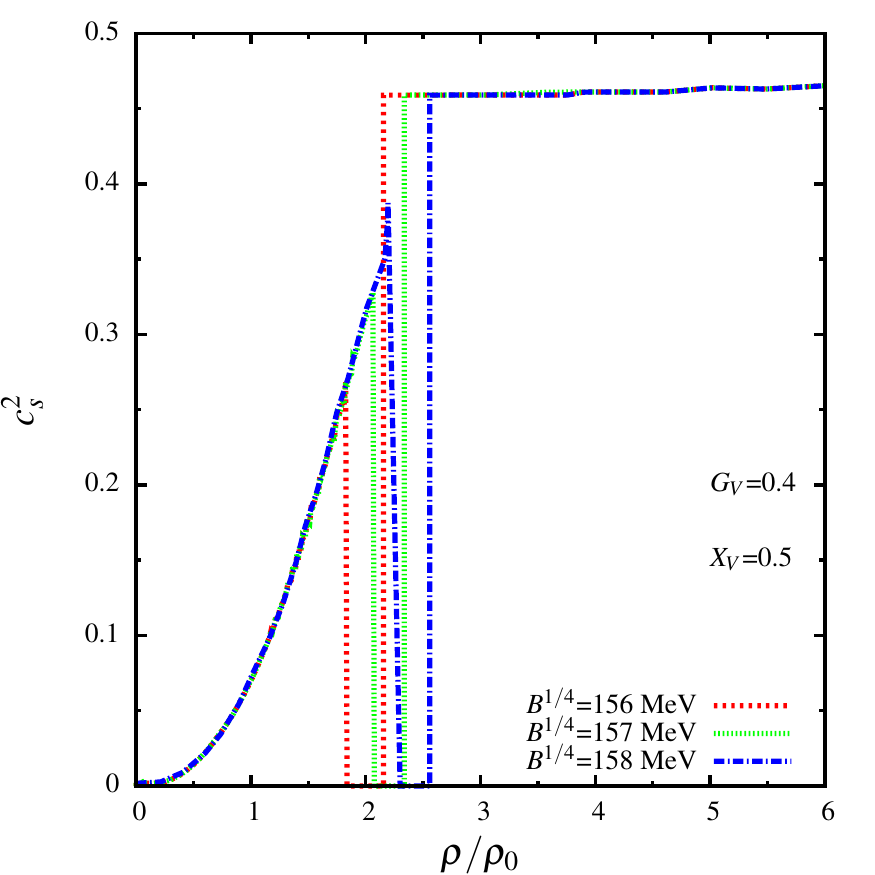}\protect\label{cs_vB}}
\hfill
\subfloat[]{\includegraphics[width=0.33\textwidth]{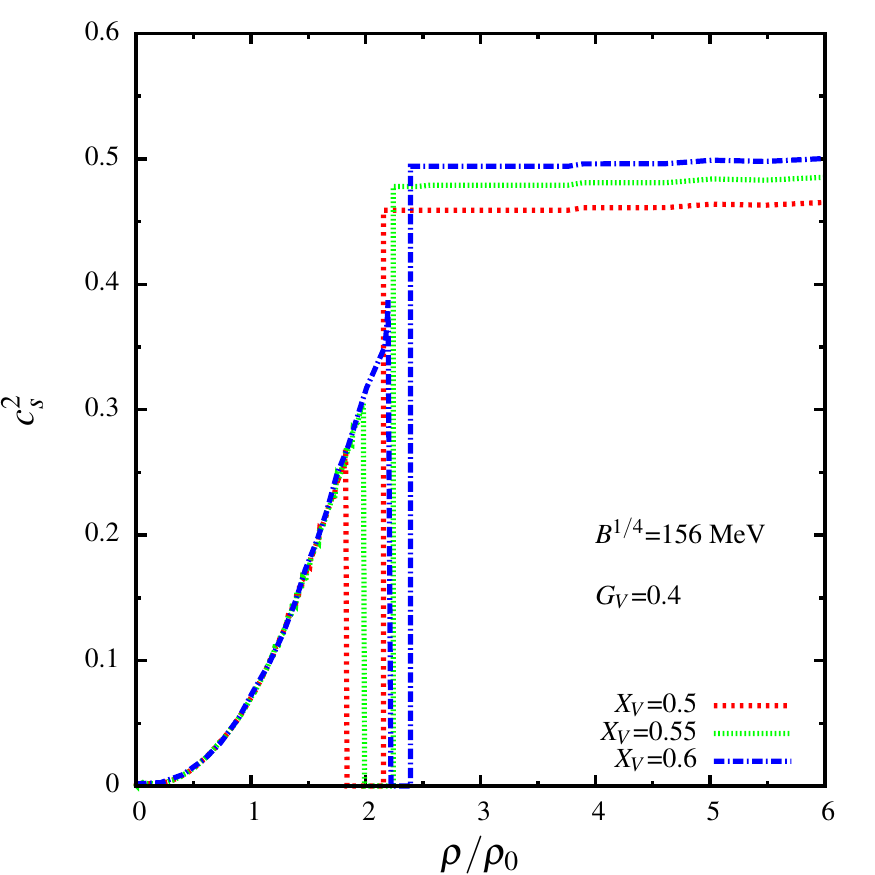}\protect\label{cs_Xv}}
\hfill
\subfloat[]{\includegraphics[width=0.33\textwidth]{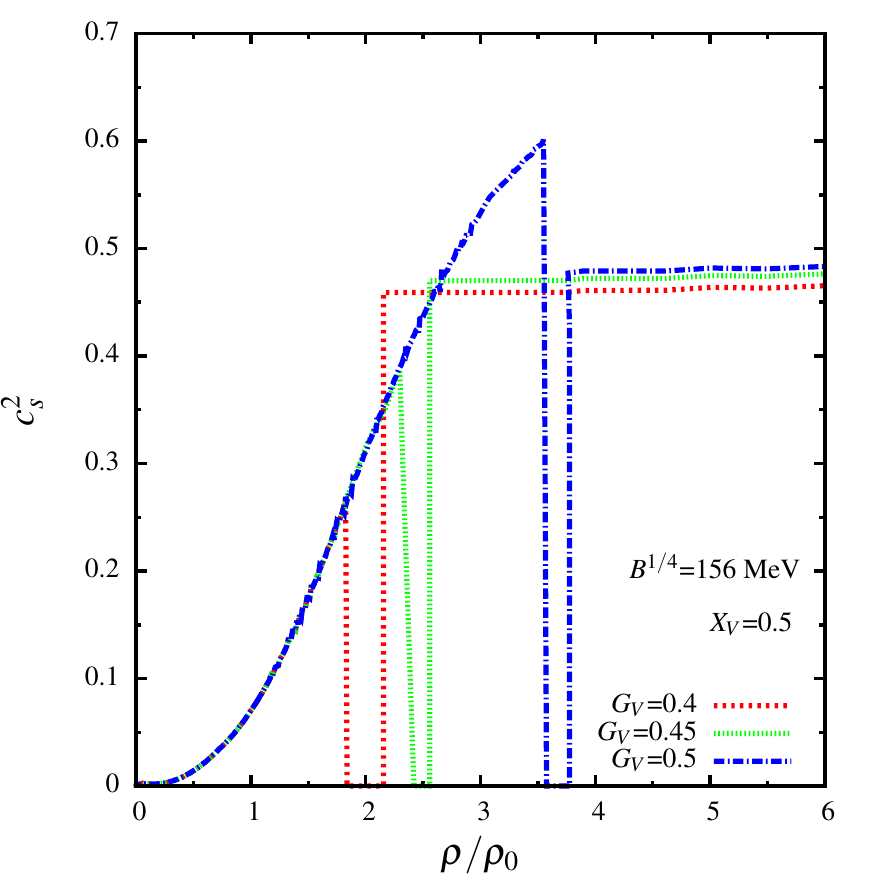}\protect\label{cs_Gv}}
\caption{\it Variation of speed of sound with density of hybrid star with vBag model for (a) different $B$ and fixed $X_V$ $b_4$, $G_V$, (b) different $X_V$ and fixed $b_4$, $G_V$, and $B$, and (c) for different $G_V$ and fixed $b_4$, $X_V$, and $B$.}
\label{cs_vBag}
\end{figure}

 Next considering the vBag model, for the variation of all the three parameters $X_V, G_V$ and $B$, as seen from figure \ref{cs_vBag}, the peak of $C_S$ is obtained mostly in the pure quark phase except for the maximum value of $G_V$ (as seen from figure \ref{cs_Gv}) for which the transition density is quite high. Similar to the previous model, we notice a flat behavior of $C_S$ in the pure quark phase. However, unlike the previous model, the increase in $C_S$ in the pure quark phase in case of the vBag model is noticeable mainly for the variation in $X_V$ as seen from figure \ref{cs_Xv}. For the variation in $G_V$ it is feeble (figure \ref{cs_Gv}) while for $B$ there is no change at all in the quark phase (figure \ref{cs_vB}) similar to that in the previous quark model (figure \ref{cs_mBag}). The maximum value (0.62) of the peak of $C_S$ in the HS matter with the vBag model is obtained for the maximum value of $G_V$ in the hadronic phase as seen from figure \ref{cs_Gv}. With this vBag model the peak of $C_S$ is obtained at values moderately higher than the conformal limit but well below the causality limit.

\begin{figure}[!ht]
\centering
\subfloat[]{\includegraphics[width=0.33\textwidth]{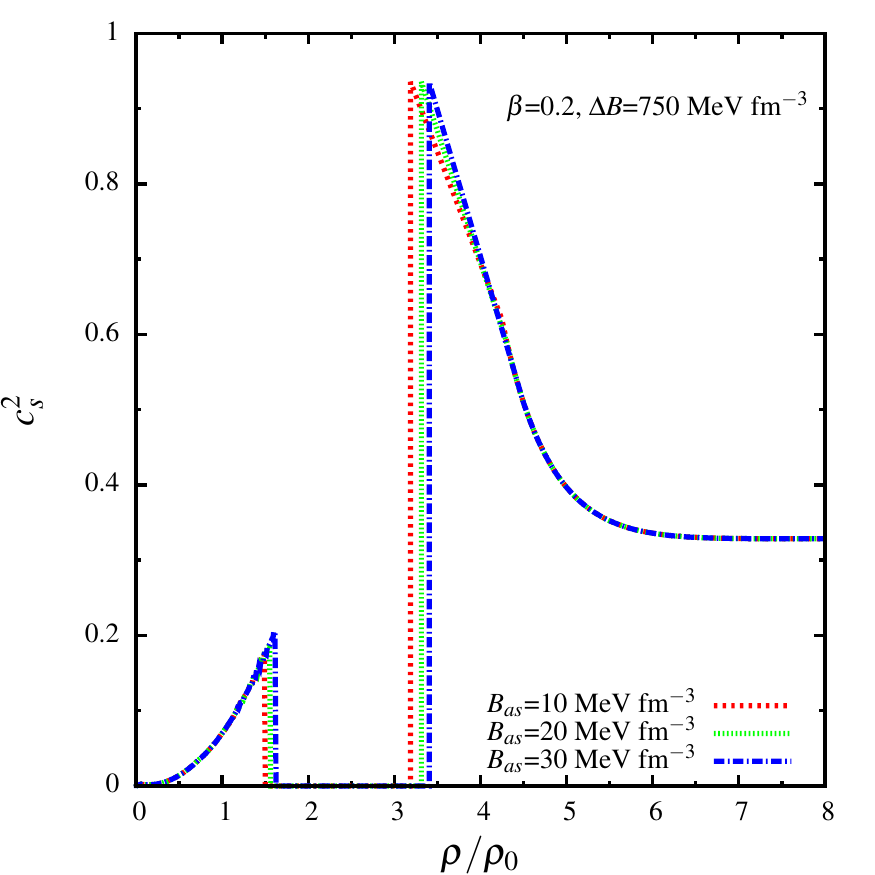}\protect\label{cs_Bas}}
\hfill
\subfloat[]{\includegraphics[width=0.33\textwidth]{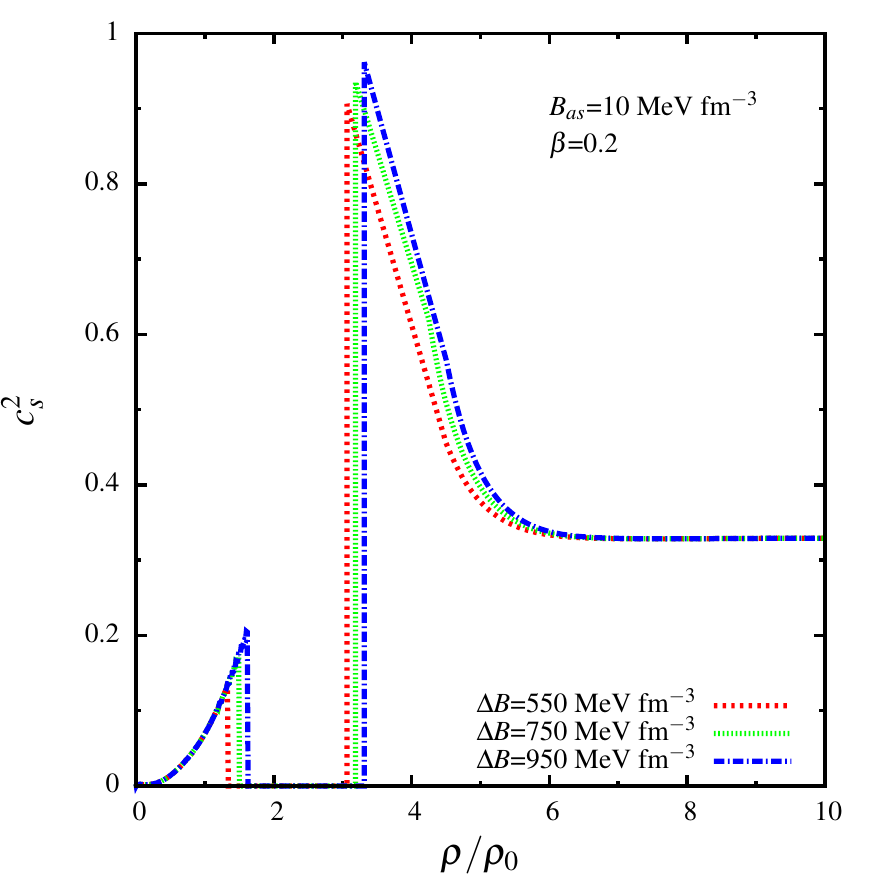}\protect\label{cs_delB}}
\hfill
\subfloat[]{\includegraphics[width=0.33\textwidth]{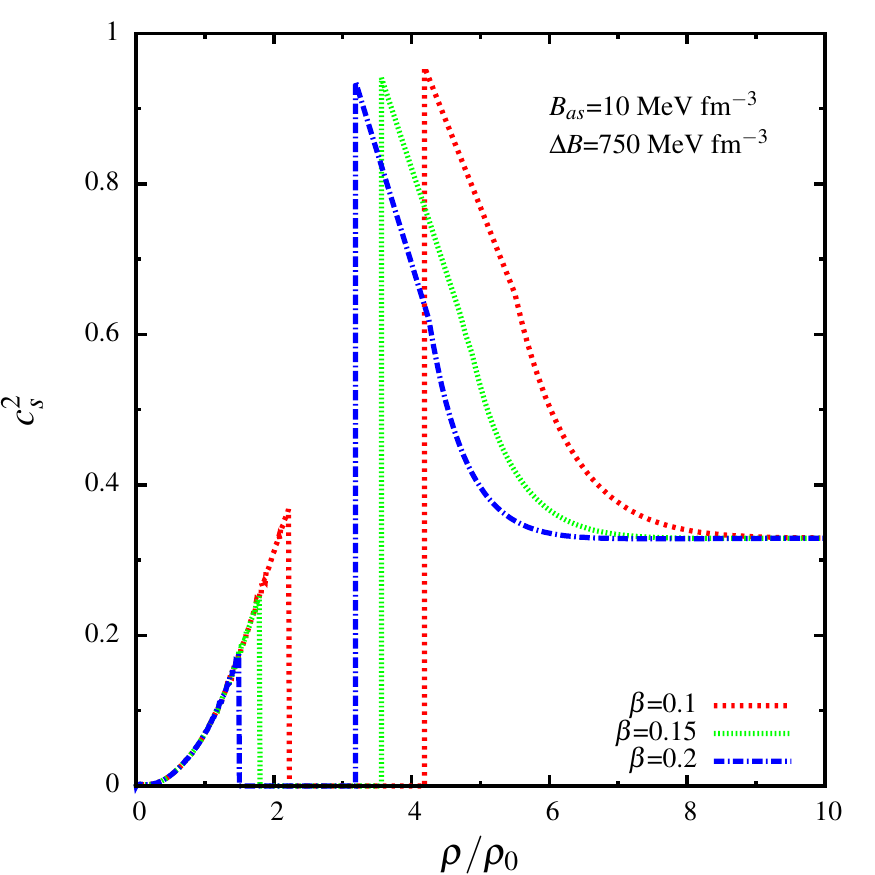}\protect\label{cs_beta}}
\caption{\it Variation of speed of sound with density of hybrid star with density dependent bag pressure in Gaussian from for (a) different $B_{as}$ and fixed $\Delta B$ and $\beta$, different $\Delta B$ and fixed $B_{as}$ and $\beta$ and (c) different $\beta$ and fixed $B_{as}$ and $\Delta B$.}
\label{cs_GDDBag}
\end{figure}

 Finally, we compare the variation of $C_S$ of HSs with the density dependent bag pressure in Gaussian form by varying $B_{as}$, $\Delta B$ and $\beta$ in figure \ref{cs_GDDBag}. The variation of all the three parameters yield very high peak values (average 0.95) of $C_S$ and for each case the peaks lie in the quark phase. The variation of $B_{as}$ shows very feeble change in $C_S$ in terms of both transition density and location of the peak as seen from figure \ref{cs_Bas}. For the variation of $\Delta B$, the change in transition density is quite feeble but the peak values of $C_S$ increases noticeably with $\Delta B$ as seen from figure \ref{cs_delB}. The variation in $\beta$ shifts considerably the transition density but the peak of $C_S$ do not show any significant change as seen from figure \ref{cs_beta}. With this model, $C_S$ peaks in the pure quark phase. For the variation of all the three parameters in this model the peak value of $C_S$ is much higher (average 0.95) than the conformal limit and is maximum compared to the results obtained with the previous quark models. It is, however, below the causality limit of speed of sound. We also find that unlike any other previous quark models, there is a steep decrease of $C_S$ up to certain values of baryon density $\rho$ depending on the chosen value of the parameters, after which $C_S$ shows no change with respect to $\rho$. The constant value that $C_S$ attains thereafter depends on the chosen value of $B_{as}$ when the quarks obtain asymptotic freedom. 
 
 Comparing the variation of $C_S$ in HSs obtained by varying the different parameters of each quark model, we find from figure \ref{cs_alpha} that the value of $C_S$ is most sensitive to the repulsive interaction parameter $\alpha$ of the first quark model considered viz. the modified bag model with strong repulsive interaction while $C_S$ is most insensitive to $m_s$. The peak value of $C_S$ is maximum in case of the quark model with density dependent bag pressure in Gaussian form. The location of the peak of $C_S$ in HSs in general also depends on the transition density.

\subsection{Hybrid Star Structure with different Quark Models}

 We now calculate the structural properties like gravitational mass ($M$), radius ($R$) and tidal deformability ($\Lambda$) of HSs using each quark model and varying their individual parameters. The obtained results are also compared with the various astrophysical constraints obtained from the different observational perspectives. We also intend to investigate particularly the parameters (of the different quark models) responsible for the formation of SPs in the $M-R$ diagram of the HSs.

\begin{figure}[!ht]
\centering
\subfloat[]{\includegraphics[width=0.33\textwidth]{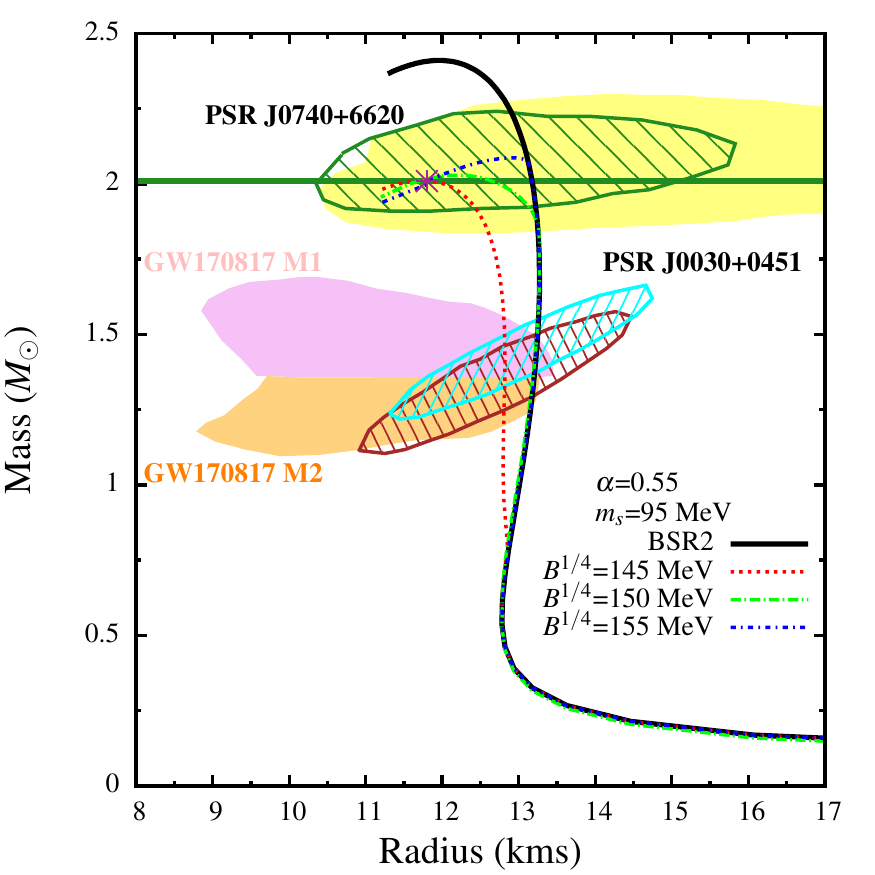}\protect\label{mr_B}}
\subfloat[]{\includegraphics[width=0.33\textwidth]{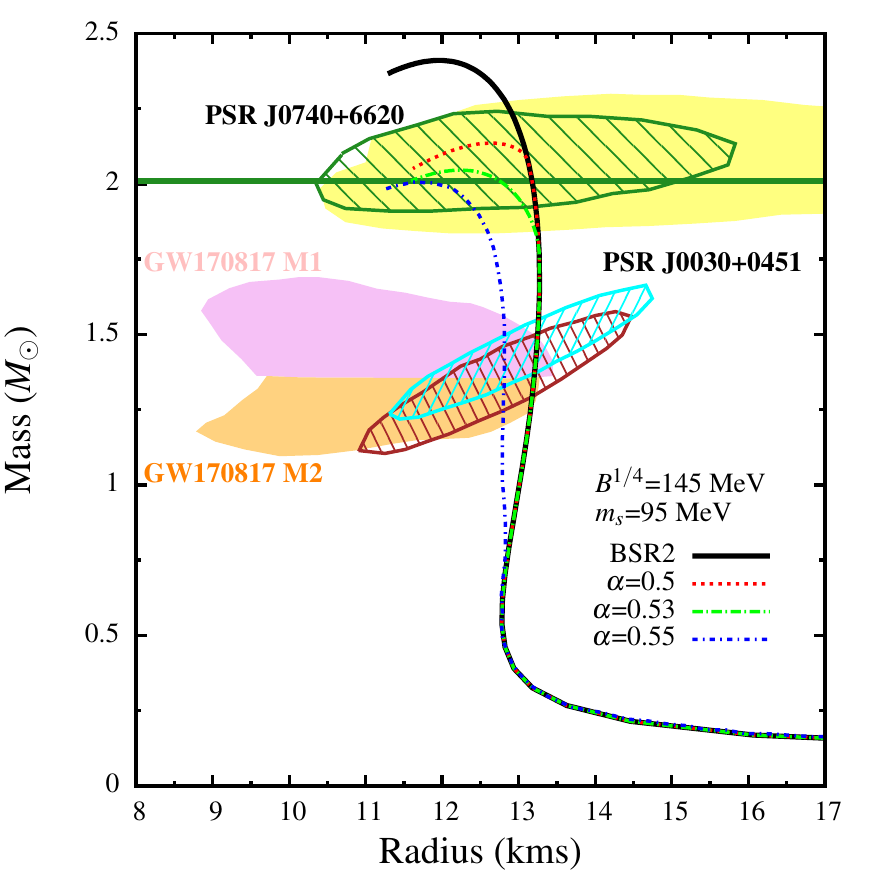}\protect\label{mr_alpha}}
\hfill
\subfloat[]{\includegraphics[width=0.33\textwidth]{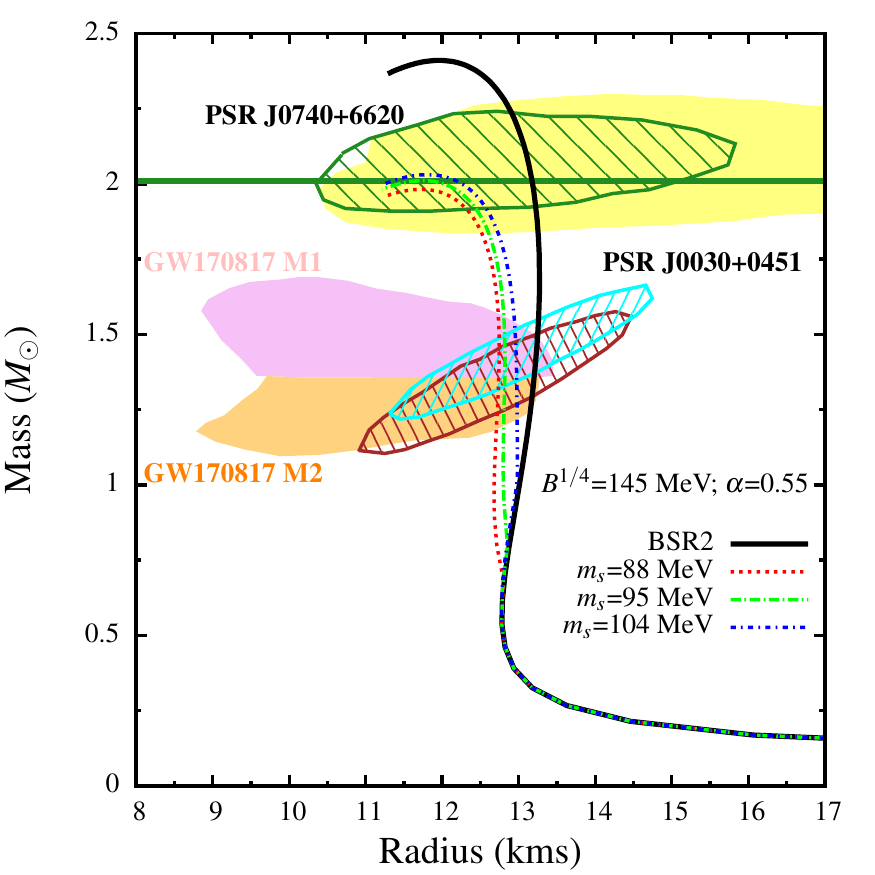}\protect\label{mr_ms}}
\caption{\it Variation of mass with radius of hybrid star with modified bag model for (a) different $B$ and fixed $\alpha$ and $m_s$ (b) different $\alpha$ and fixed $B$ and $m_s$ (c) different $m_s$ and fixed $B$ and $\alpha$. Observational limits imposed from the most massive pulsar PSR J0740+6620 ($M = 2.08 \pm 0.07 M_{\odot}$) \cite{Fonseca2021} and $R = 13.7^{+2.6}_{-1.5}$ km \cite{Miller2021} or $R = 12.39^{+1.30}_{-0.98}$ km \cite{Riley2021}) are also indicated. The constraints on $M-R$ plane prescribed from GW170817 \cite{GW170817}) and NICER experiment for PSR J0030+0451 \cite{Riley2019,Miller2019} are also compared. The position of special point is marked with asterisks.}
\label{mr_mBag}
\end{figure}

\begin{figure}[!ht]
\centering
\subfloat[]{\includegraphics[width=0.33\textwidth]{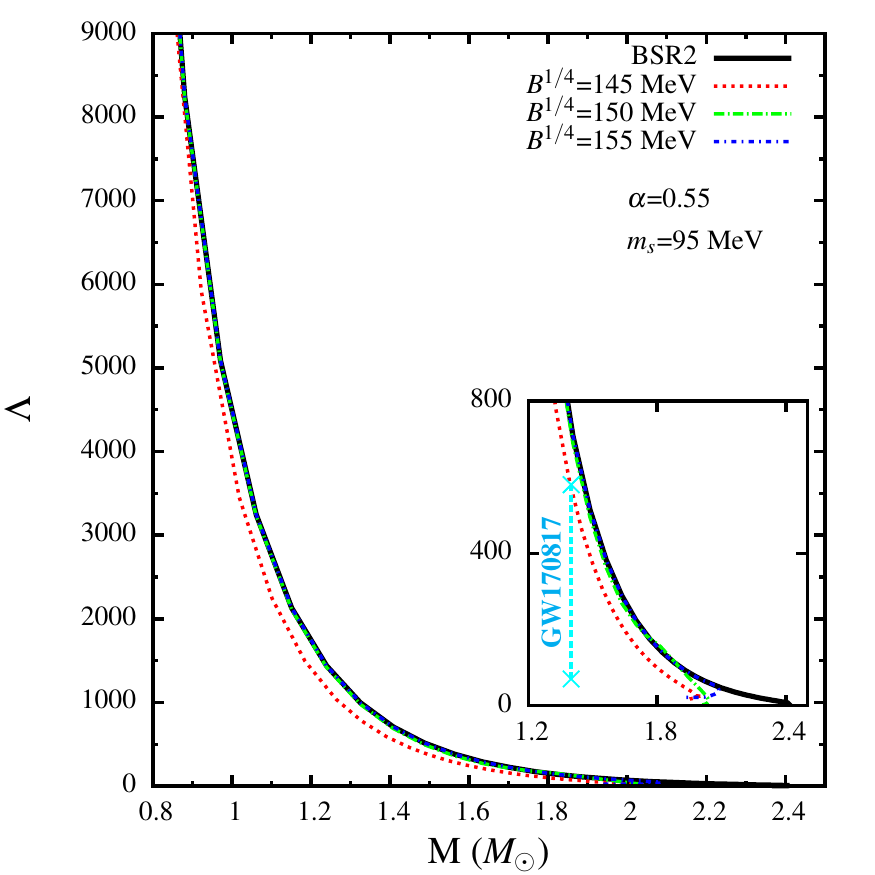}\protect\label{LamM_B}}
\hfill
\subfloat[]{\includegraphics[width=0.33\textwidth]{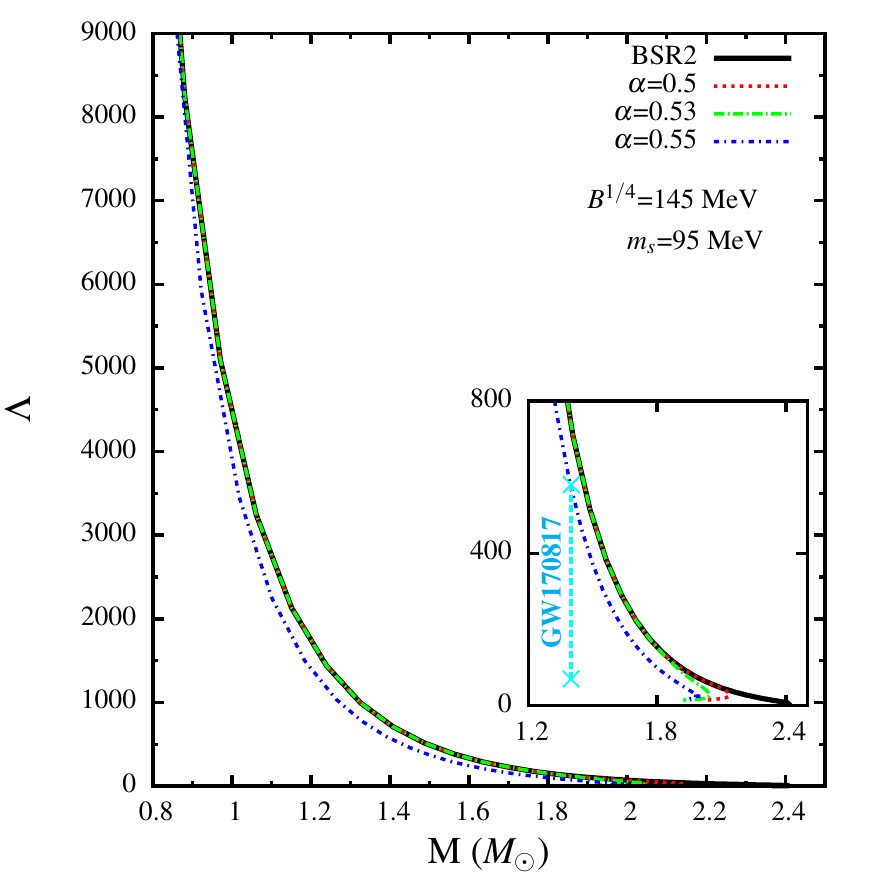}\protect\label{LamM_alpha}}
\hfill
\subfloat[]{\includegraphics[width=0.33\textwidth]{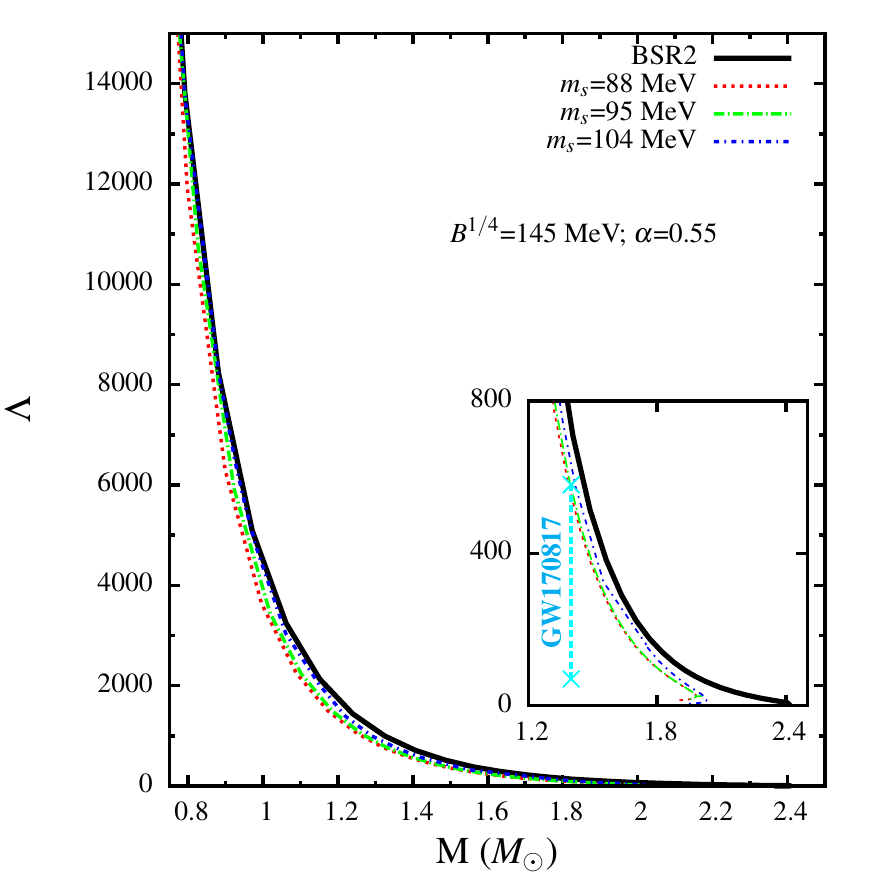}\protect\label{LamM_ms}}
\caption{\it Corresponding variation of tidal deformability with mass. The constraint on $\Lambda_{1.4}$ from GW170817 \cite{GW170817} is also shown.}
\label{mLam_mBag}
\end{figure}

 In figure \ref{mr_mBag} we show the variation of mass with radius of HSs with the different parameters of the first quark model i.e. the modified bag model with strong repulsive interaction. Comparing figures \ref{mr_B}, \ref{mr_alpha} and \ref{mr_ms} the $M-R$ variation is most sensitive to the repulsive interaction parameter $\alpha$. The maximum mass of HS ($M_{max}$) is specially affected by slight variation of $\alpha$ as seen from figure \ref{mr_alpha}. On the other hand $B$ has moderate effect on $M_{max}$ while $m_s$ the least and almost insignificant similar to the case of $C_S$ as seen in figure \ref{cs_ms}. $M_{max}$ is higher for high values of both $B$ and $m_s$ and low values of $\alpha$. The maximum mass and radius of the HSs with this model are obtained for the minimum value of $\alpha$=0.5 as seen from figure \ref{mr_alpha}. For the different choice of $B$, $\alpha$ and $m_s$, the HS configurations satisfy the different astrophysical constraints from different observations such as the mass-radius values of PSR J0740+6620 and PSR J0030+0451 obtained from NICER experiment and that from GW170817 data. We also notice that the variation of $B$ leads to the emergence of a distinct special point (SP) (marked with asterisk - ($M_{SP}$, $R_{SP}$)), irrespective of the value of $B$ or the transition density on the $M-R$ plot \ref{mr_B}. The SP in this case is located on the unstable branch of the HSs formed after the maximum value of $M$. We note that the variation of the other two parameters viz. $\alpha$ and $m_s$ of this quark model do not show the formation of SPs. In figure \ref{mLam_mBag} we also study the effect of tidal deformability with respect to mass for the variation of $B$, $\alpha$ and $m_s$. The junction correction in the context of calculating the value of $k_2$ \cite{k2_corr} do not bring any perceptible change in the net value of $\Lambda$ except at the hadron-quark interface where we obtain smoother transitions in terms of $\Lambda$. In the main figures of \ref{mLam_mBag} the transition points can also be noticed in terms of $\Lambda$ while from the insets it is seen that the constraint on $\Lambda_{1.4}$ from GW170817 data is better satisfied with lower values of $B$ and higher values of $\alpha$. The values of $\Lambda_{1.4}$ obtained within the range $m_s$=95$^{+11}_{-5}$ MeV are consistent with that obtained from GW170817. The combination of the three parameters ($B,\alpha,m_s$) that satisfy all the present astrophysical constraints the best is (145 MeV, 0.55, 95 MeV).

\begin{figure}[!ht]
\centering
\subfloat[]{\includegraphics[width=0.33\textwidth]{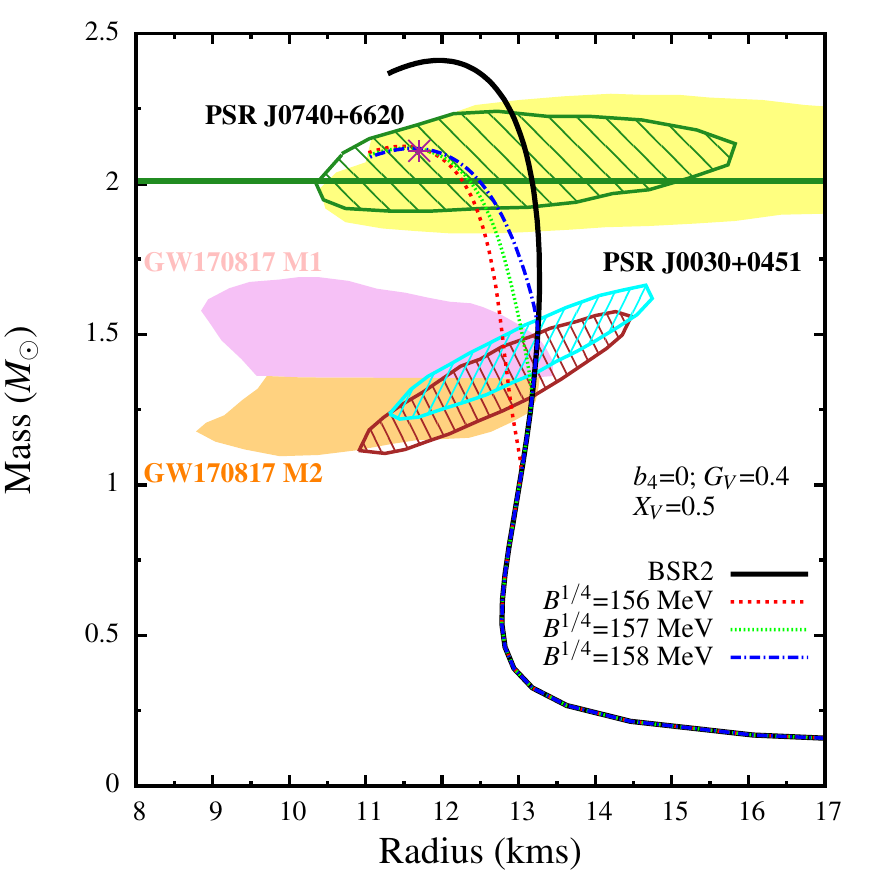}\protect\label{mr_B2}}
\subfloat[]{\includegraphics[width=0.33\textwidth]{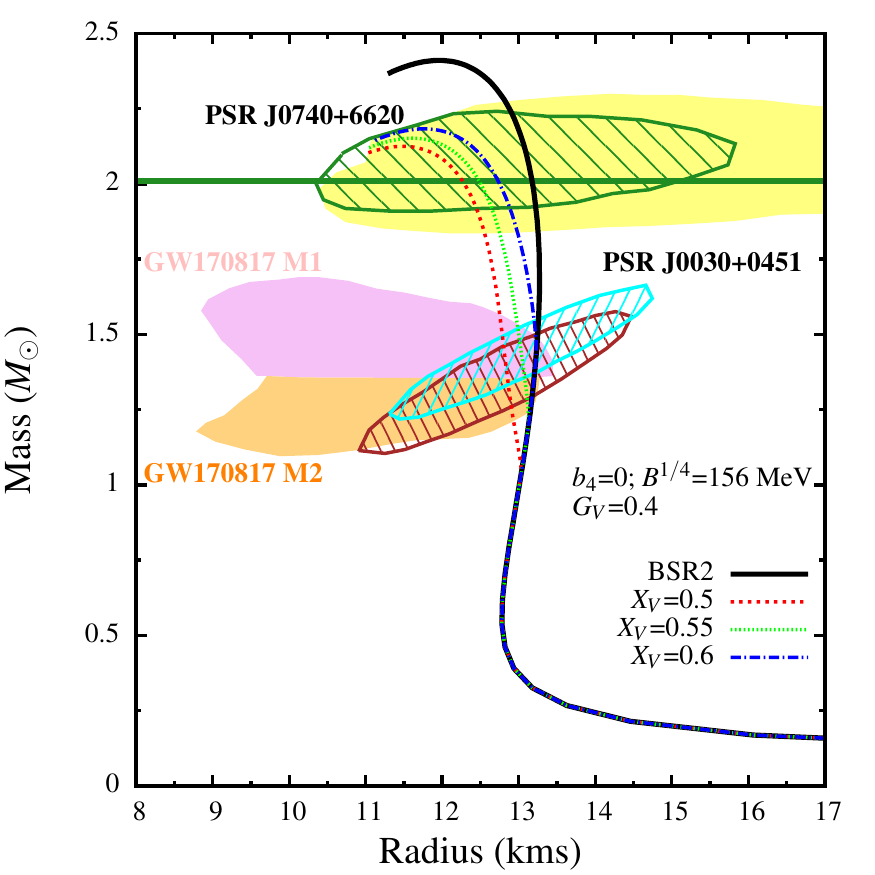}\protect\label{mr_Xv}}
\hfill
\subfloat[]{\includegraphics[width=0.33\textwidth]{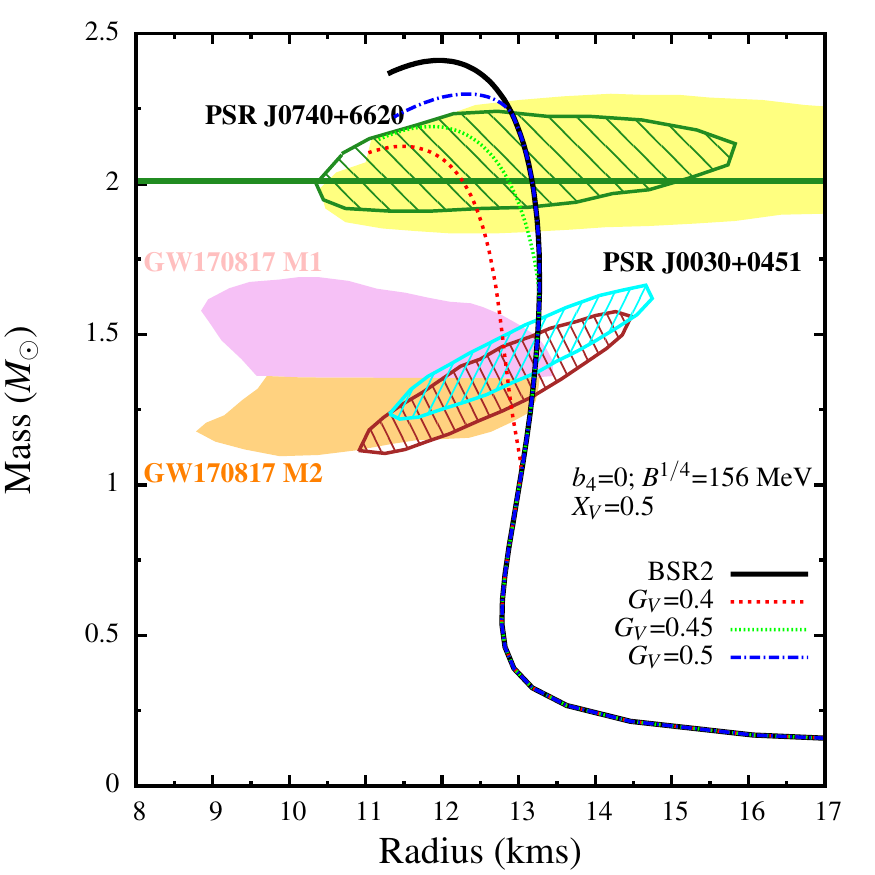}\protect\label{mr_Gv}}
\caption{\it Variation of mass with radius of hybrid star with vBag model for (a) different $B$ and fixed $X_V$ $b_4$, $G_V$ (b) different $X_V$ and fixed $b_4$, $G_V$ and $B$ (c) different $G_V$ and fixed $b_4$, $X_V$ and $B$.}
\label{mr_vBag}
\end{figure}

\begin{figure}[!ht]
\centering
\subfloat[]{\includegraphics[width=0.33\textwidth]{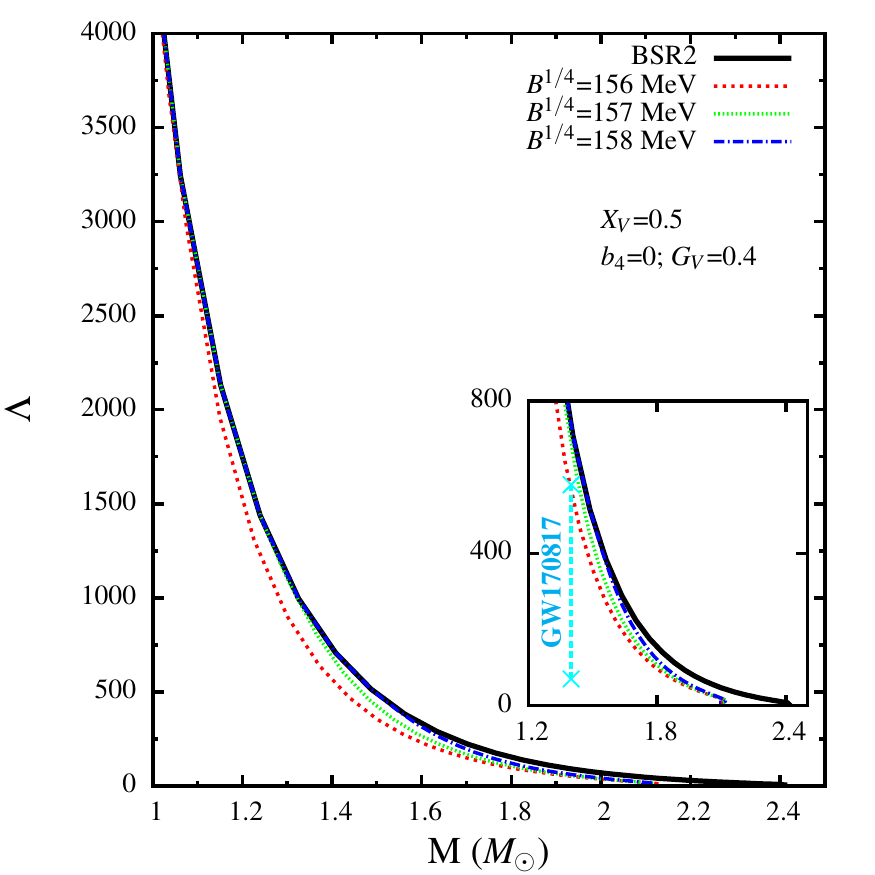}\protect\label{LamM_B2}}
\hfill
\subfloat[]{\includegraphics[width=0.33\textwidth]{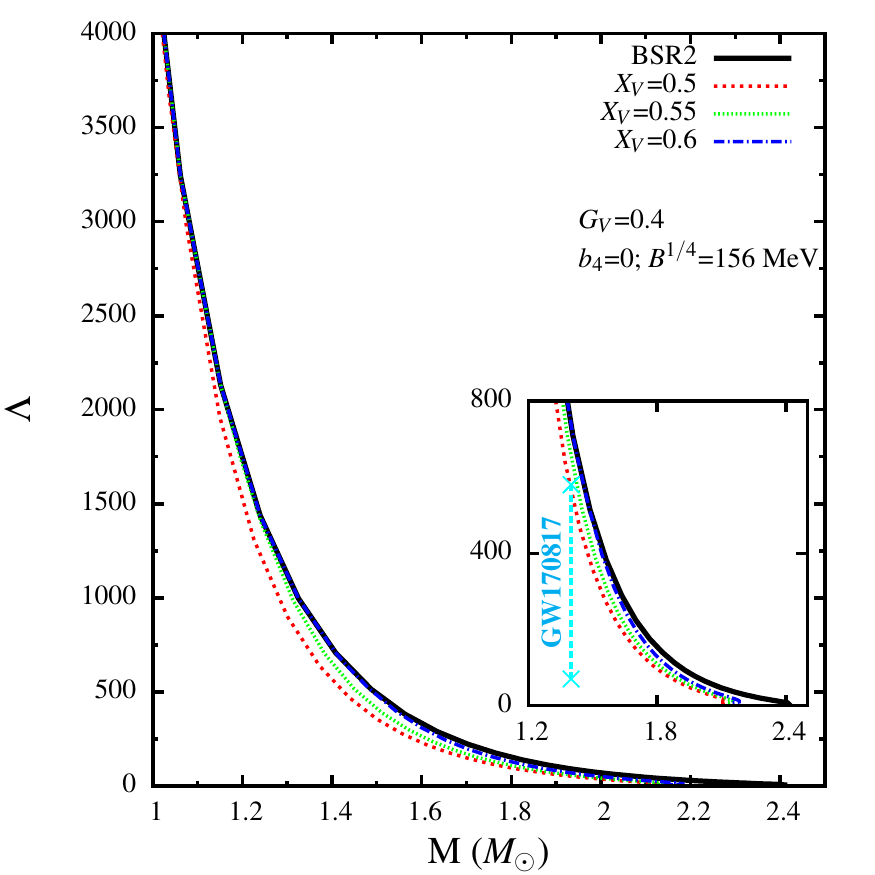}\protect\label{LamM_Xv}}
\hfill
\subfloat[]{\includegraphics[width=0.33\textwidth]{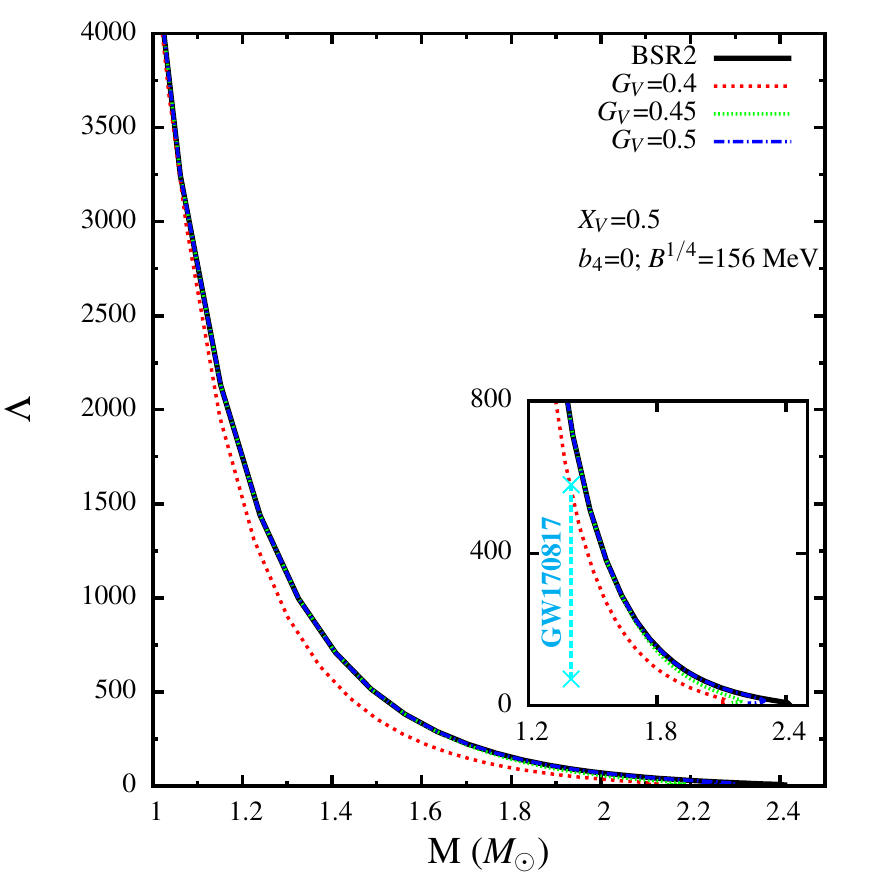}\protect\label{LamM_Gv}}
\caption{\it Corresponding variation of tidal deformability with mass.}
\label{mLam_vBag}
\end{figure}

 We next display the variation of mass with radius of HSs with the different parameters of the second quark model i.e. the vBag model in figure \ref{mr_vBag}. The maximum mass of HS increases with both $G_V$ and $X_V$ while the reverse trend is noticed for variation of $B$ since there is a very feeble increase in $M_{max}$ for decreasing values of $B$ as seen from figure \ref{mr_B2}. The maximum mass and radius of the HSs with vBag model are obtained for the maximum value of $G_V$=0.5 as seen from figure \ref{mr_Gv}. The maximum mass and radius of HSs are also most sensitive to this parameter $G_V$ for this quark model. The variation of $X_V$ also shows a moderate effect on $M_{max}$ as seen from figure \ref{mr_Xv} unlike the effect of variation of $B$ as seen from \ref{mr_B2}. The parameters of this model also have good impact on the value of $R_{1.4}$ of the HSs specially for the variation of $B$ and $X_V$ compared to the variation of $G_V$ since transition is earlier in the case of variation of $B$ and $X_V$ compared to that of $G_V$. Similar to the HSs obtained with the previous quark model as shown in figure \ref{mr_mBag}, we find that even for this vBag model, the only parameter responsible for the formation of SPs is the bag pressure $B$ (as seen from figure \ref{mr_B2}). However, for the vBag model the location of SP (as marked with asterisks in figure \ref{mr_B2}) is on the stable second branch of HSs unlike that in figure \ref{mr_B}. The variation of $G_V$ or $X_V$ do not yield SPs on the $M-R$ variation of HSs. For the different choice of $B$, $G_V$ and $X_V$, the HS configurations satisfy the different astrophysical constraints from different observations as discussed before. In figure \ref{mLam_vBag} we also study the corresponding effect of tidal deformability with respect to mass for the variation of $B$, $G_V$ and $X_V$. The constraint on $\Lambda_{1.4}$ from GW170817 data is well satisfied only with the minimum values of $B$, $G_V$ and $X_V$ as seen from the insets. For this model the combination of the three parameters ($B,X_V,G_V$) that satisfy all the present astrophysical constraints the best is (156 MeV, 0.5, 0.4).

\begin{figure}[!ht]
\centering
\subfloat[]{\includegraphics[width=0.33\textwidth]{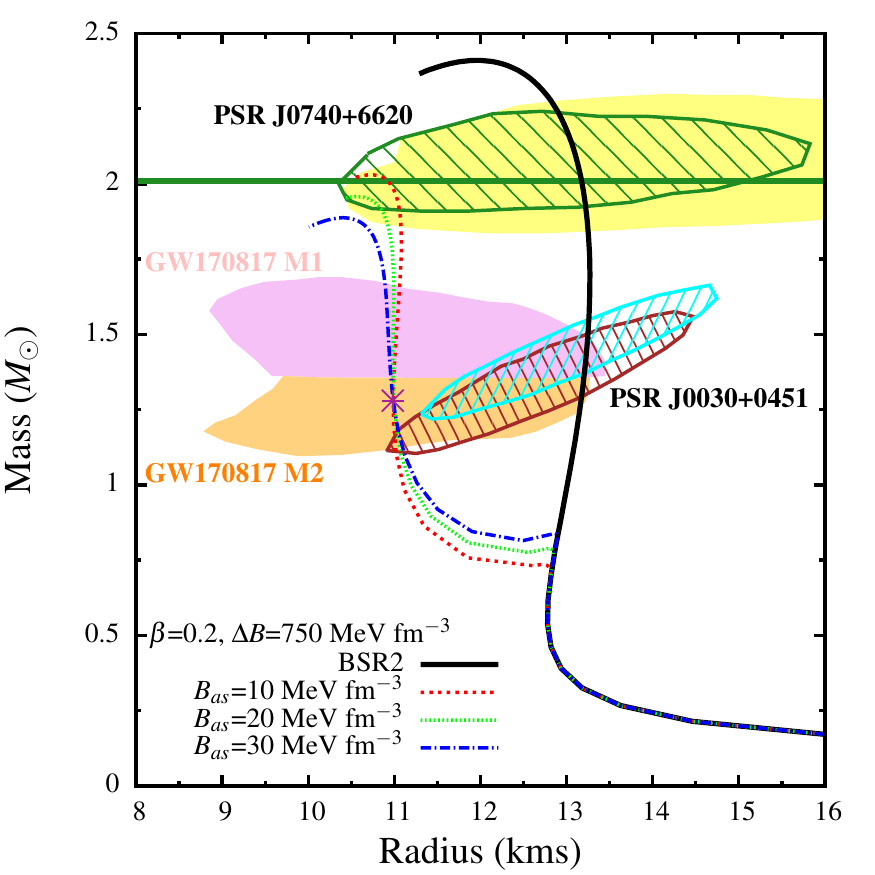}\protect\label{mr_Bas}}
\subfloat[]{\includegraphics[width=0.33\textwidth]{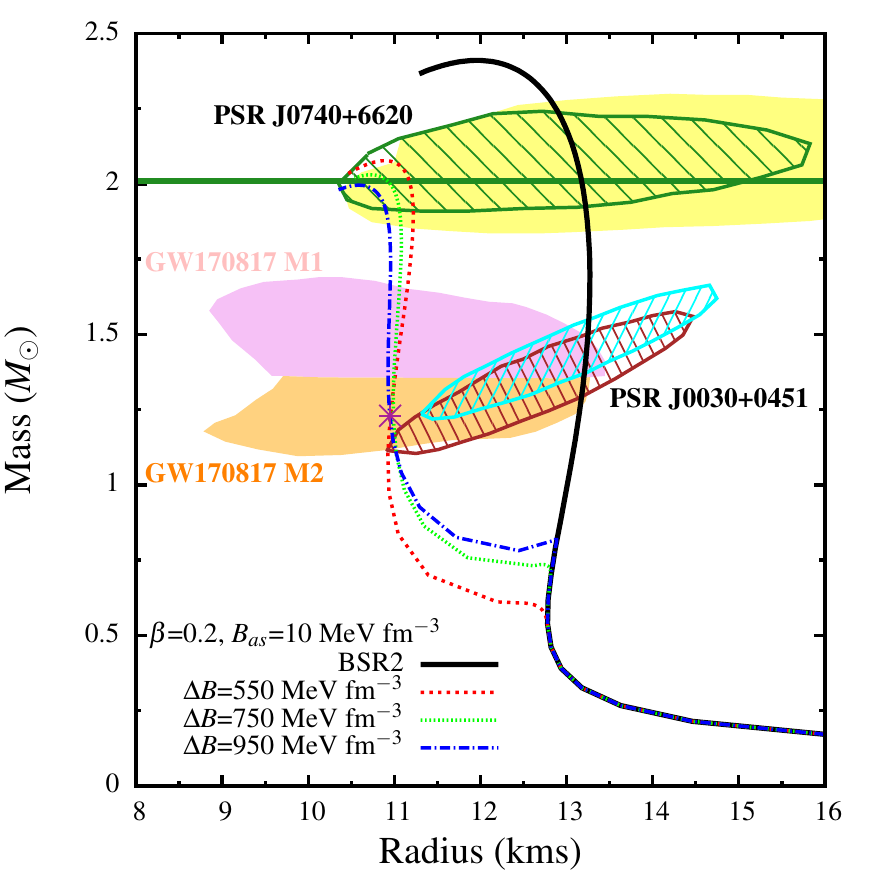}\protect\label{mr_delB}}
\hfill
\subfloat[]{\includegraphics[width=0.33\textwidth]{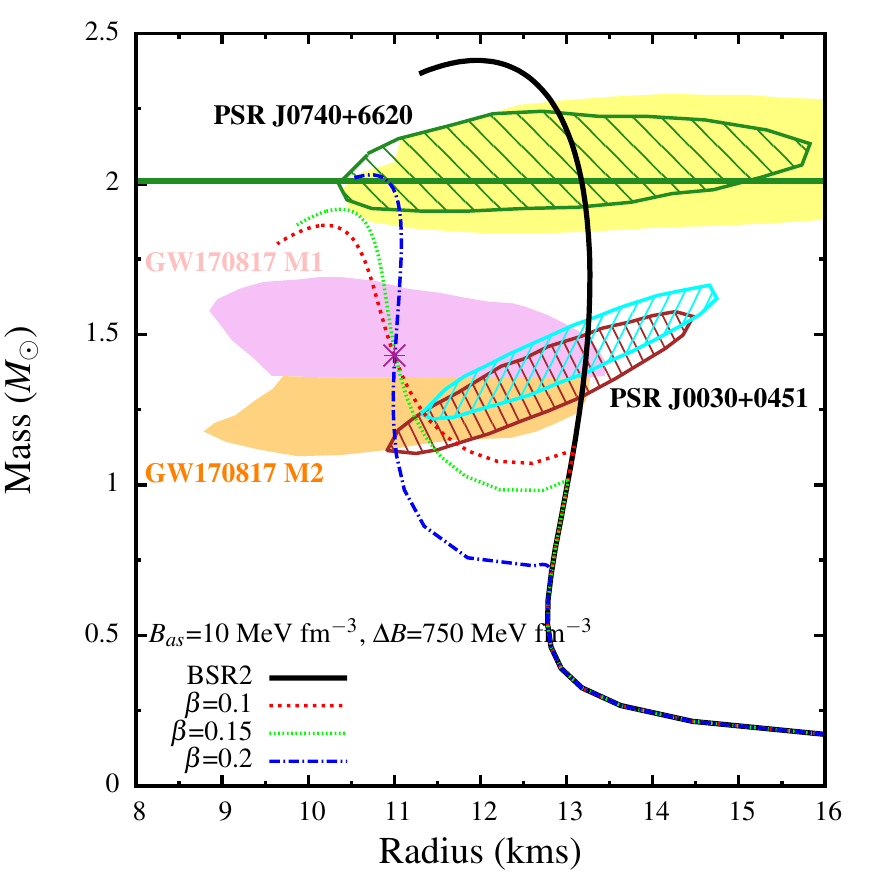}\protect\label{mr_beta}}
\caption{\it Variation of mass with radius of hybrid star with density of hybrid star with density dependent bag pressure in Gaussian from for (a) different $B_{as}$ and fixed $\Delta B$, and $\beta$ (b) different $\Delta B$ and fixed $B_{as}$, and $\beta$ (c) different $\beta$ and fixed $B_{as}$, and $\Delta B$.}
\label{mr_GDDBag}
\end{figure}

\begin{figure}[!ht]
\centering
\subfloat[]{\includegraphics[width=0.33\textwidth]{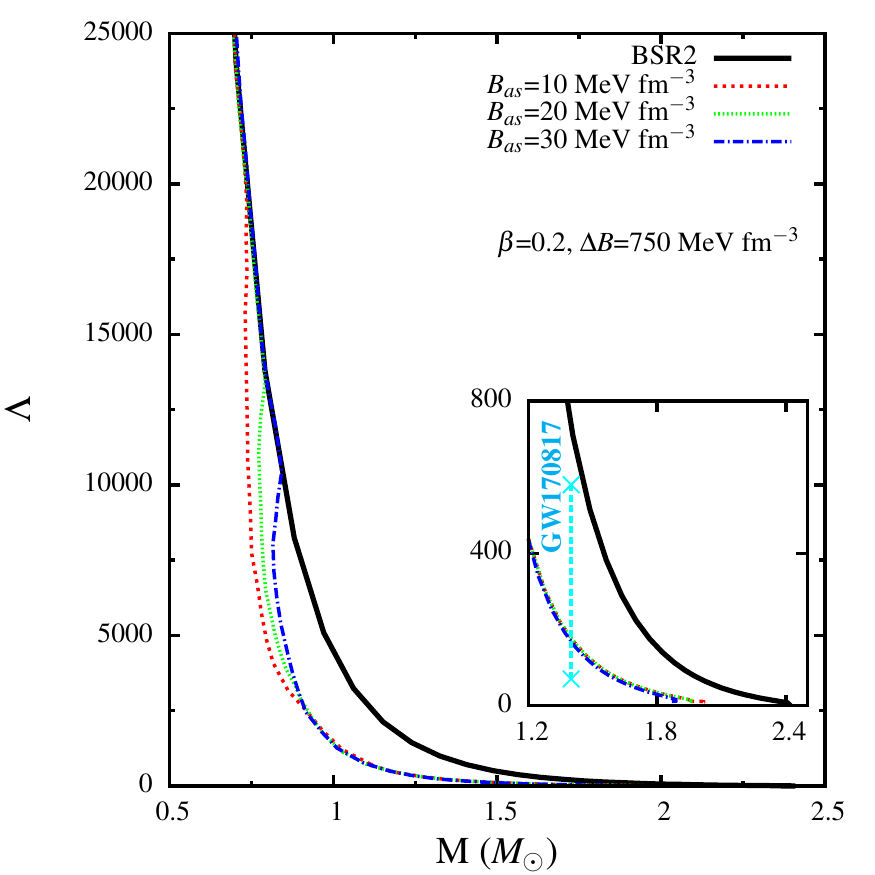}\protect\label{LamM_Bas}}
\hfill
\subfloat[]{\includegraphics[width=0.33\textwidth]{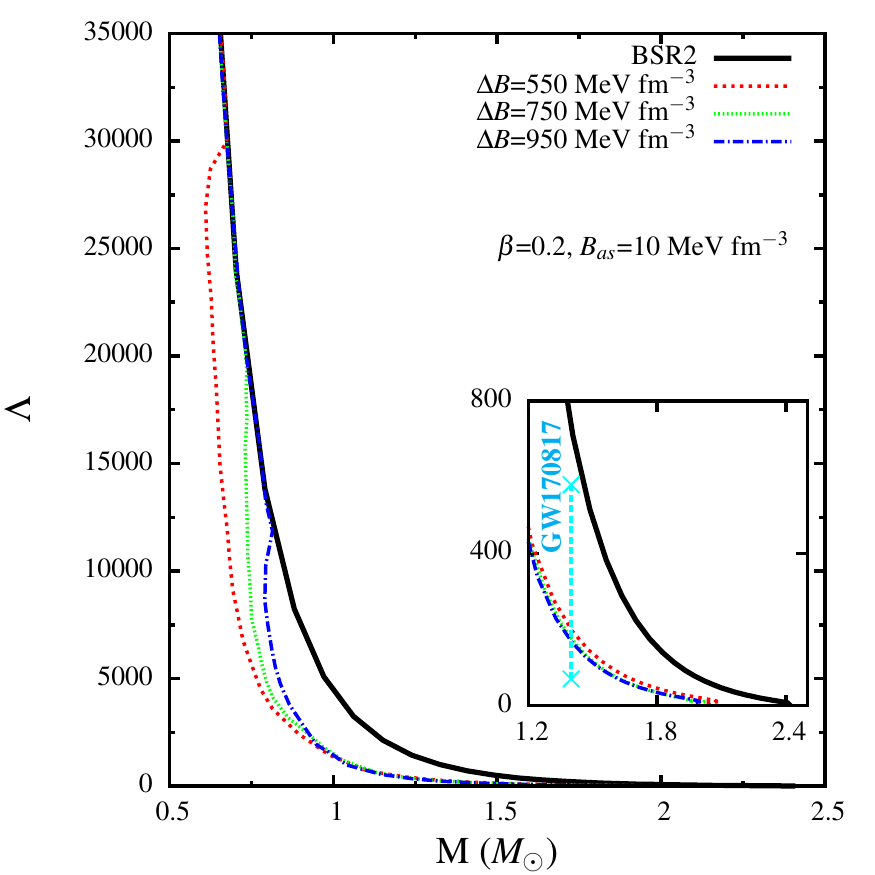}\protect\label{LamM_delB}}
\hfill
\subfloat[]{\includegraphics[width=0.33\textwidth]{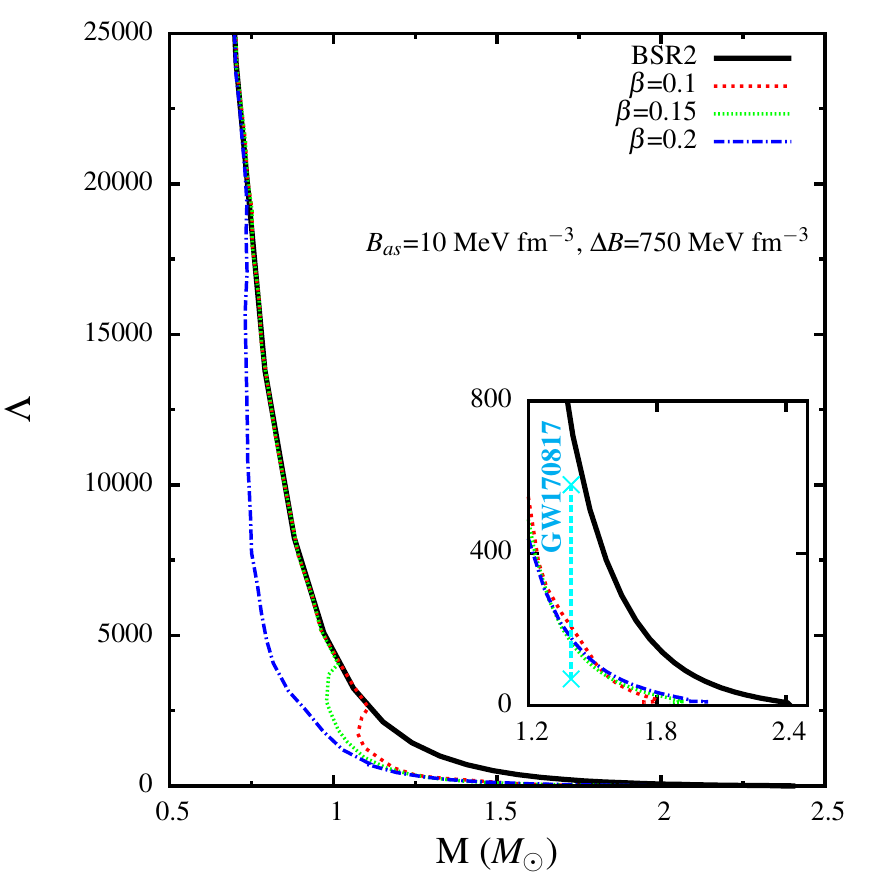}\protect\label{LamM_beta}}
\caption{\it Corresponding variation of tidal deformability with mass.}
\label{mLam_GDDBag}
\end{figure}

 In figure \ref{mr_GDDBag} we present the variation of mass with radius of HSs with the different parameters of the third quark model i.e. density dependence of bag pressure in a Gaussian form. Most of the HS configurations obtained with this model exhibit twin star characteristics. The maximum mass increases with increasing $\Delta B$ and $\beta$ but decreases with increasing $B_{as}$. The maximum mass and radius of the HSs with this model are obtained for the minimum value of $\Delta B$=550 MeV fm$^{-3}$ as seen from figure \ref{mr_delB}. With this model, $M_{max}$ and $R_{1.4}$ of the HSs are most sensitive to the value of $\beta$. Overall, the HS configurations obtained with this quark model satisfy the various astrophysical constraints on compact star structural properties from various perspectives except for $B_{as}>$ 10 MeV fm$^{-3}$ and $\beta<$ 0.2 for which the constraint on $M_{max}$ from PSR J0740+6620 is not satisfied. Interestingly, we notice the formation of distinct SPs due to the variation of all the three parameters ($B_{as}$, $\Delta B$ and $\beta$) involved in this form of the bag model. Such SPs are marked with asterisks in figures \ref{mr_Bas}, \ref{mr_delB} and \ref{mr_beta}. It is also seen the position of the SPs are also quite close for the three cases of variation of $B_{as}$, $\Delta B$ and $\beta$. This once again imply that the bag pressure plays a crucial role in the emergence of SPs on the mass-radius diagram of HSs. In figure \ref{mLam_GDDBag} we also study the corresponding effect of tidal deformability with respect to mass for the variation of $B_{as}$, $\Delta B$ and $\beta$. The insets show that the constraint on $\Lambda_{1.4}$ from GW170817 data is very well satisfied for all the chosen values of $B_{as}$, $\Delta B$ and $\beta$. For this model the combination of the three parameters ($B_{as}$, $\Delta B$, $\beta$) that satisfy all the present astrophysical constraints the best are (10 MeV fm$^{-3}$, 550 MeV fm$^{-3}$, 0.2), (10 MeV fm$^{-3}$, 750 MeV fm$^{-3}$, 0.2) and (10 MeV fm$^{-3}$, 950 MeV fm$^{-3}$, 0.2).

\begin{figure}[!ht]
\centering
\subfloat[]{\includegraphics[width=0.33\textwidth]{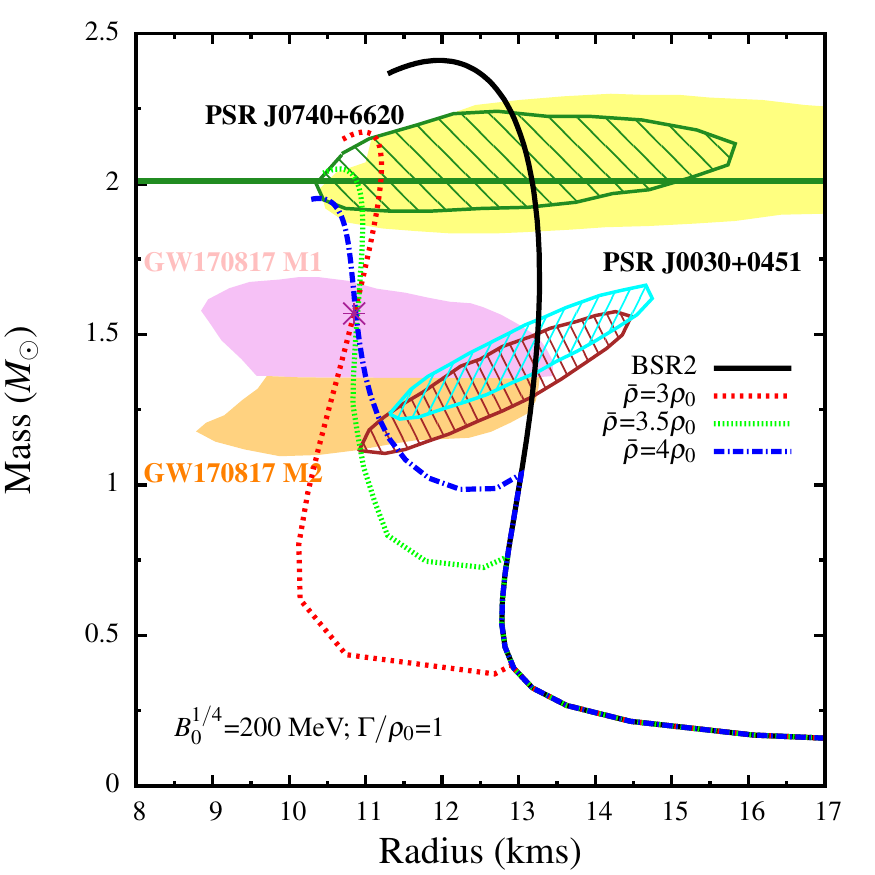}\protect\label{mr_rhobar}}
\subfloat[]{\includegraphics[width=0.33\textwidth]{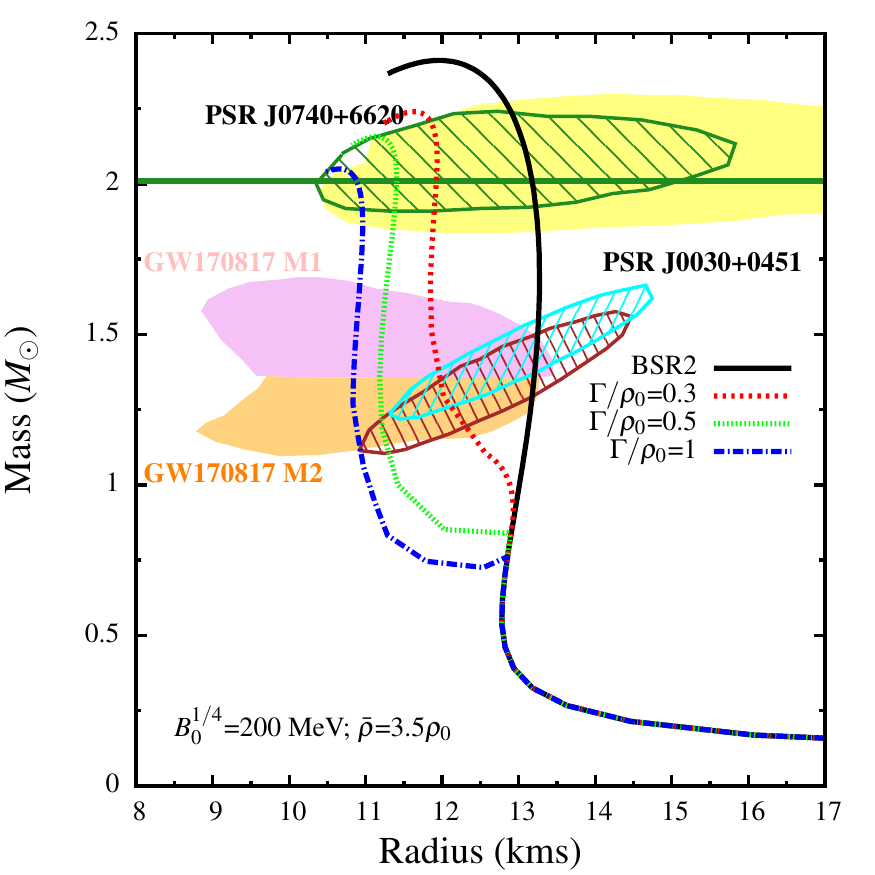}\protect\label{mr_gamma}}
\hfill
\subfloat[]{\includegraphics[width=0.33\textwidth]{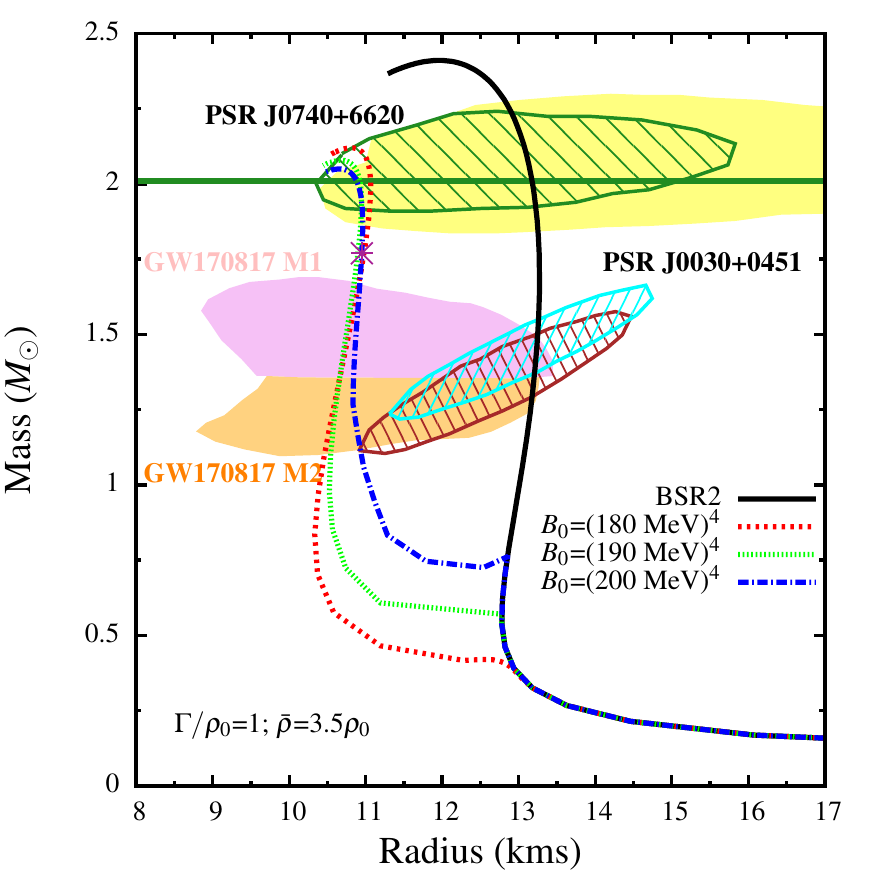}\protect\label{mr_B0}}
\caption{\it Variation of mass with radius of hybrid star with density dependent bag pressure in hyperbolic from for (a) different $\bar{\rho}$ and fixed $B_0$ and $\Gamma/\rho_0$ (b) different $\Gamma/\rho_0$ and fixed $B_0$ and $\bar{\rho}$ (c)different $B_0$ and fixed $\Gamma/\rho_0$ and $\bar{\rho}$.}
\label{mr_HDDBag}
\end{figure}

\begin{figure}[!ht]
\centering
\subfloat[]{\includegraphics[width=0.33\textwidth]{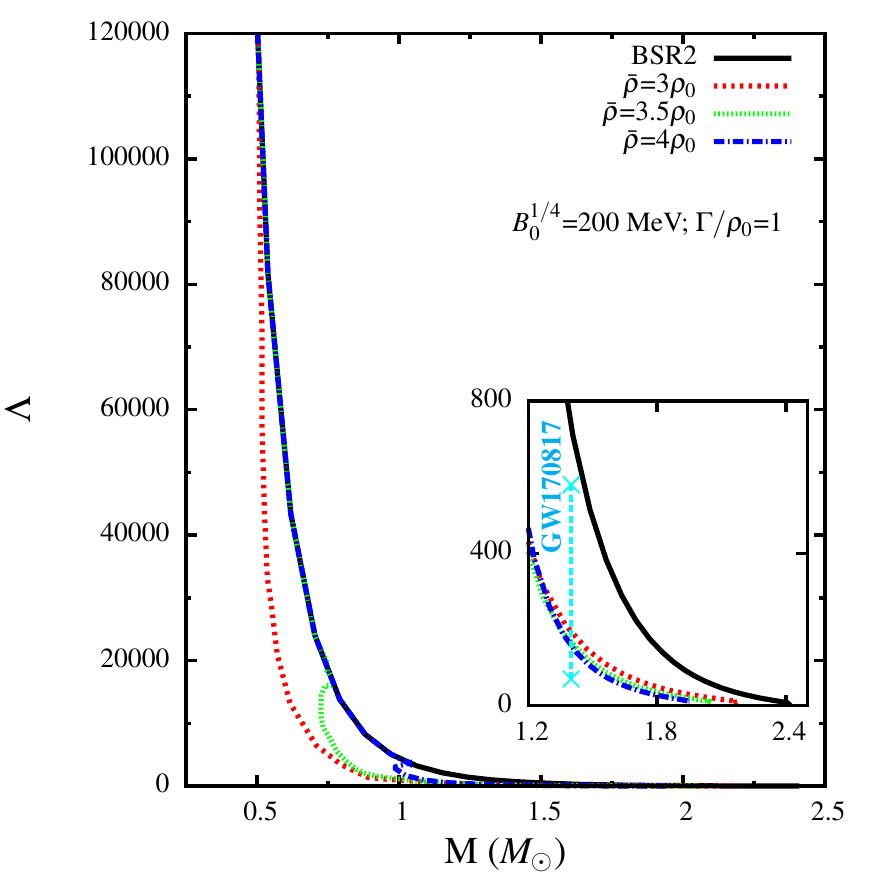}\protect\label{LamM_rhobar}}
\hfill
\subfloat[]{\includegraphics[width=0.33\textwidth]{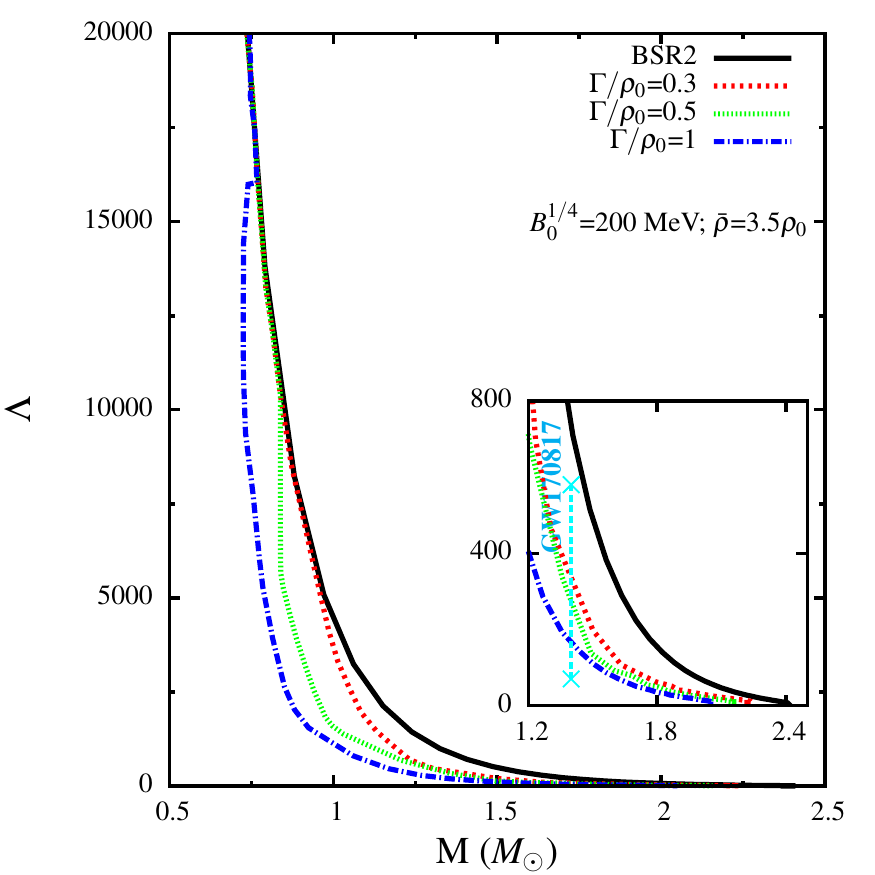}\protect\label{LamM_gamma}}
\hfill
\subfloat[]{\includegraphics[width=0.33\textwidth]{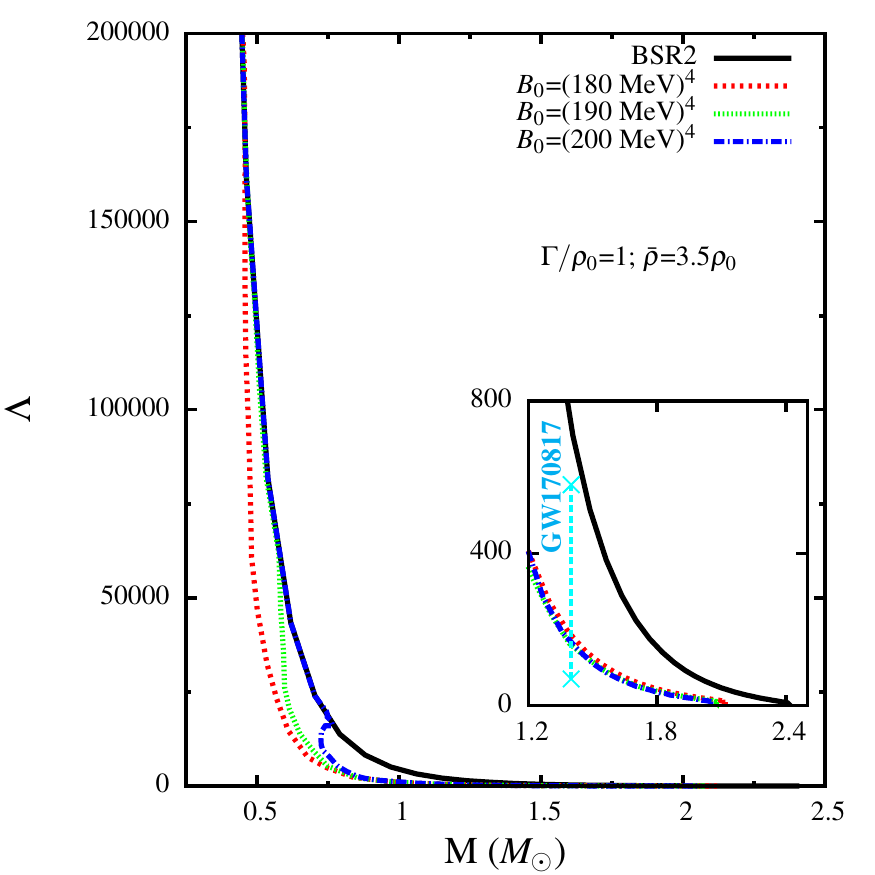}\protect\label{LamM_B0}}
\caption{\it Corresponding variation tidal deformability with mass.}
\label{mLam_HDDBag}
\end{figure}

 Finally, in figure \ref{mr_HDDBag} we present the variation of mass with radius of HSs with the different parameters of the fourth and last quark model considered in this work i.e. density dependence of bag pressure in a hyperbolic form. Similar to the HS configurations obtained with the previous quark model with density dependent bag pressure in Gaussian form, we notice twin star characteristics in most of the $M-R$ curves of HSs obtained with this quark model where the hyperbolic form of the $B(\rho)$ dependence is considered. The maximum mass increases with decreasing values of $\bar{\rho}$, $\Gamma/\rho_0$ and $B_0$. The maximum mass and radius of the HSs with this model are obtained for the minimum value of $\Gamma/\rho_0$ as seen from figure \ref{mr_gamma}. With this model, $M_{max}$ and $R_{1.4}$ of the HSs are quite sensitive to all the three parameters. All the HS configurations obtained with this quark model satisfy the maximum mass constraint from PSR J0740+6620 except for the maximum value of $\bar{\rho}$ as seen from figure \ref{mr_rhobar}. The constraint from GW170817 data is well satisfied by all the HS configurations obtained with this quark model. We notice from figures \ref{mr_rhobar}, \ref{mr_gamma} and \ref{mr_B0} that the NICER data for PSR J0030+0451 is better satisfied with increasing values of all the three parameters individually. Like the HSs with density dependent bag pressure with Gaussian from, we also obtain SPs for HSs with density dependent bag pressure in hyperbolic from. However, for the later we obtain SPs only for the variations of $\bar{\rho}$ and $B_0$ (figures \ref{mr_rhobar} and \ref{mr_B0}) and not $\Gamma/\rho_0$ (figure \ref{mr_gamma}). In figure \ref{mLam_HDDBag} we also study the corresponding effect of tidal deformability with respect to mass for the variation of $\bar{\rho}$, $\Gamma/\rho_0$ and $B_0$. From the insets we find that the constraint on $\Lambda_{1.4}$ from GW170817 data is well satisfied for the chosen values of $\bar{\rho}$, $\Gamma/\rho_0$ and $B_0$.

 Considering the results of the structural properties of HSs with all the four quark models, we find that both in the density dependent and independent cases the variation of bag pressure leads to the formation of SPs on the $M-R$ diagram of HSs as seen from figures \ref{mr_B}, \ref{mr_B2}, \ref{mr_Bas}, \ref{mr_delB},\ref{mr_beta}, \ref{mr_rhobar} and \ref{mr_B0}. Interestingly, it is also seen from figures \ref{LamM_B}, \ref{LamM_B2}, \ref{LamM_Bas}, \ref{LamM_delB}, \ref{LamM_beta}, \ref{LamM_rhobar} and \ref{LamM_B0} that the curves overlap near $M_{SP}$. This indicates that the feature of SPs is also exhibited also in all the $\Lambda - M$ dependence corresponding to the $M-R$ dependence for which SPs are noticed.

\begin{figure}[!ht]
\centering
{\includegraphics[width=0.5\textwidth]{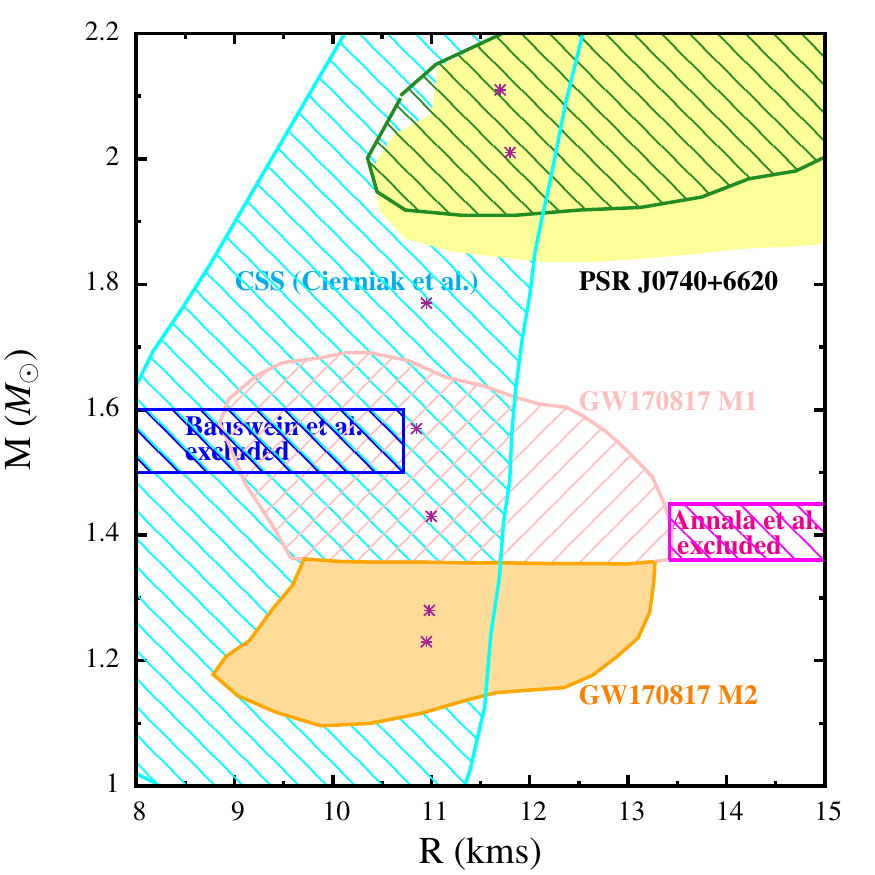}}
\caption{\it Location of special points (asterisks) on the mass-radius plot of hybrid stars for the variation of bag pressure in density dependent and independent scenarios. The possible positions of the special points for the CSS quark model with $C_S^2$ = 0.7 \cite{Cierniak} is also compared. The allowed \cite{GW170817} and excluded \cite{Bauswein2} regions on the mass-radius plane from GW170817 are also indicated.}
\label{SP}
\end{figure}

 As discussed in the Introduction section \ref{Intro}, SPs are often treated as universal properties of HSs that help us to understand the possible existence of HSs in the light of the multi-messenger data as signals \cite{Cierniak,Cierniak2,Sen9}. In figure \ref{SP} we compare our results of location of SPs with that of \cite{Cierniak} obtained with the CSS quark model. We find that our locations of the SPs with different forms of MIT Bag model are within the possible region of SPs prescribed by \cite{Cierniak} for constant speed of sound $C_S^2$ = 0.7 with the CSS quark model. Also, the excluded regions of the $M-R$ plane as prescribed from the GW170817 analysis \cite{Bauswein2} are not violated by our locations of the SPs. In two cases when the bag pressure is density independent, our ($M_{SP},R_{SP}$) values also satisfy the $M-R$ constraint from PSR J0740+6620. This is also seen from figures \ref{mr_B} and \ref{mr_B2} unlike the case when the bag pressure is density dependent for which the location of SPs do not satisfy the $M-R$ constraint from PSR J0740+6620. This was also noticed in our previous work \cite{Sen9} where this constraint was not satisfied by the locations of the SPs of HSs obtained with the quark model with density dependent bag pressure and six different hadronic models.


\section{Summary and Conclusion}
\label{Conclusion}

 We investigate the possibility of hadron-quark phase transition in compact star cores and the formation of HSs. For the purpose we employ the RMF hadronic model BSR2 while the quark phase is described by four different forms of the MIT Bag model. Phase transition is achieved with Maxwell construction and as a result sharp transition with density jumps is observed. The variation of speed of sound is studied in HS matter. The location of the peak of $C_S$ can lie in the hadronic or quark phase depending on the different parameters as well as the quark model considered and the transition density. Considering the results of HSs with all the four forms of bag model, we find that the value of $C_S$ is most sensitive to the repulsive interaction parameter $\alpha$ of the modified bag model with strong repulsive interaction and most insensitive to the mass of s quark $m_s$ while $C_S$ is maximum for the case of density dependent bag pressure in Gaussian from.
 
 We then studied the structural properties of the HSs obtained with the four different forms of the quark model for different parameters involved in these four types of quark models with respect to the present day astrophysical constraints on the $M-R$ relation obtained from PSR J0740+6620, GW170817, and NICER experiment for PSR J0030+0451 and also that on $\Lambda_{1.4}$ obtained from GW170817 data analysis. Even for $M_{max}$ and $R_{1.4}$ of the HSs, $m_s$ remains the most insensitive parameter considering all the various other parameters involved in the four different quark models. The value of $\Lambda_{1.4}$ from GW170817 data is best satisfied in the density dependent scenario of the bag pressure compared to those where it is taken to be constant. However, the NICER data for PSR J0030+0451 is better satisfied when the bag pressure is treated to be constant. It is also interesting to note that twin star characteristics is obtained in certain cases only when the bag pressure is considered to be density dependent and not when it is taken to be constant.

 Interestingly, we notice the emergence of SPs on the $M-R$ diagram of HSs and considering the $M-R$ dependence of HSs for the variation of different parameters involved in the four different forms of MIT Bag model, we find that the bag pressure plays immense role in the emergence of SPs in both density dependent and independent scenarios. No other parameter associated with the four quark models can lead to the formation of SPs. This is the most important finding of the present work. Even in the $\Lambda - M$ dependence corresponding to the $M-R$ diagrams for which SPs are noticed, the feature of SPs is also well prominent. With respect to the different constraints on the $M-R$ plain of HSs, we also found that the SPs lie on the allowed regions and their locations do not violate the GW170817 excluded zones. The ($M_{SP},R_{SP}$) values satisfy the $M-R$ constraint from PSR J0740+6620 only when the bag pressure is taken to be constant.  


\section*{Acknowledgement}

The authors thank Dr. Naosad Alam, TIFR Mumbai for providing the hadronic EoS and also for useful discussions.

\end{document}